\documentclass[sigconf,nonacm,numbers]{acmart}

\usepackage{color}
\usepackage{adjustbox}
\usepackage{subcaption}
\usepackage{algorithm}
\usepackage{algpseudocode}
\usepackage{multirow}
\usepackage{hyperref}
\usepackage{booktabs}
\usepackage{amsmath}

\usepackage{amsfonts}
\usepackage{amssymb}
\usepackage{balance}
\usepackage{array}
\newcolumntype{C}[1]{>{\centering\arraybackslash}p{#1}}

\AtBeginDocument{%
  }

\copyrightyear{2024}
\acmYear{2024}
\setcopyright{rightsretained}
\acmConference[IMC '24]{Proceedings of the 2024 ACM Internet Measurement Conference}{November 4--6, 2024}{Madrid, Spain}
\acmBooktitle{Proceedings of the 2024 ACM Internet Measurement Conference (IMC '24), November 4--6, 2024, Madrid, Spain}\acmDOI{10.1145/3646547.3688437}
\acmISBN{979-8-4007-0592-2/24/11}

\usepackage{listings, xcolor}

\definecolor{verylightgray}{rgb}{.97,.97,.97}

\lstdefinelanguage{Solidity}{
	keywords=[1]{anonymous, assembly, assert, balance, break, call, callcode, case, catch, class, constant, continue, contract, debugger, default, delegatecall, delete, do, else, emit, event, export, external, false, finally, for, function, constructor, gas, if, implements, import, in, indexed, instanceof, interface, internal, is, length, library, log0, log1, log2, log3, log4, memory, modifier, new, payable, pragma, private, protected, public, pure, push, require, return, returns, revert, selfdestruct, send, storage, struct, suicide, super, switch, then, this, throw, transfer, true, try, typeof, using, value, view, while, with, addmod, ecrecover, keccak256, mulmod, ripemd160, sha256, sha3}, 
	keywordstyle=[1]\color{blue}\bfseries,
	keywords=[2]{address, bool, byte, bytes, bytes1, bytes2, bytes3, bytes4, bytes5, bytes6, bytes7, bytes8, bytes9, bytes10, bytes11, bytes12, bytes13, bytes14, bytes15, bytes16, bytes17, bytes18, bytes19, bytes20, bytes21, bytes22, bytes23, bytes24, bytes25, bytes26, bytes27, bytes28, bytes29, bytes30, bytes31, bytes32, enum, int, int8, int16, int24, int32, int40, int48, int56, int64, int72, int80, int88, int96, int104, int112, int120, int128, int136, int144, int152, int160, int168, int176, int184, int192, int200, int208, int216, int224, int232, int240, int248, int256, mapping, string, uint, uint8, uint16, uint24, uint32, uint40, uint48, uint56, uint64, uint72, uint80, uint88, uint96, uint104, uint112, uint120, uint128, uint136, uint144, uint152, uint160, uint168, uint176, uint184, uint192, uint200, uint208, uint216, uint224, uint232, uint240, uint248, uint256, var, void, ether, finney, szabo, wei, days, hours, minutes, seconds, weeks, years},	
	keywordstyle=[2]\color{teal}\bfseries,
	keywords=[3]{block, blockhash, acoinbase, difficulty, gaslimit, number, timestamp, msg, data, gas, sender, sig, value, now, tx, gasprice, origin},	
	keywordstyle=[3]\color{violet}\bfseries,
	identifierstyle=\color{black},
	sensitive=false,
	comment=[l]{//},
	morecomment=[s]{/*}{*/},
	commentstyle=\color{gray}\ttfamily,
	stringstyle=\color{red}\ttfamily,
	morestring=[b]',
	morestring=[b]"
}

\usepackage{listings, xcolor}

\definecolor{verylightgray}{rgb}{.97,.97,.97}

\definecolor{darkyellow}{rgb}{.55,.40,.0}

\definecolor{olive}{rgb}{.45,.60,.15}

\definecolor{niceblue}{rgb}{.15,.30,.60}

\lstdefinelanguage{Cypher}{
	keywords=[1]{match, return, where}, 
	keywordstyle=[1]\color{olive}\bfseries,
	keywords=[2]{not, exists, in, and, or},	
	keywordstyle=[2]\color{niceblue}\bfseries,
	identifierstyle=\color{black},
	sensitive=false,
	comment=[l]{//},
	morecomment=[s]{/*}{*/},
	commentstyle=\color{gray}\ttfamily,
	stringstyle=\color{darkyellow}\ttfamily,
	morestring=[b]',
	morestring=[b]"
}

\newcommand{\PaperTitle}{Analyzing the Impact of Copying-and-Pasting Vulnerable Solidity Code Snippets from Question-and-Answer Websites}

\newcommand{\toolname}{\textsc{CCD}}

\begin{document}

\title{\PaperTitle}

\author{Konrad Weiss}
\affiliation{%
  \institution{Fraunhofer AISEC}
  \city{Garching near Munich}
  \country{Germany}}
\email{konrad.weiss@aisec.fraunhofer.de}

\author{Christof Ferreira Torres}
\affiliation{%
  \institution{ETH Zurich}
  \city{Zurich}
  \country{Switzerland}}
\email{christof.torres@inf.ethz.ch}

\author{Florian Wendland}
\affiliation{%
  \institution{Fraunhofer AISEC}
  \city{Garching near Munich}
  \country{Germany}}
\email{florian.wendland@aisec.fraunhofer.de}


\newcommand{\vulnsnipp} {4,596}
\newcommand{\analyzedsnippets} {18,660}
\newcommand{\snippetswclones} {723}
\newcommand{\dissemsnipwc} {602}
\newcommand{\sourcesnipwc} {227}
\newcommand{\clonesofdissemsnip} {26,565}
\newcommand{\clonesofsourcesnip} {5,395}
\newcommand{\unqclonesofdissemsnip} {24,451}
\newcommand{\unqclonesofsourcesnip} {4,828}
\newcommand{\nrvalidationinfpath} {19,992}
\newcommand{\nrvalidationiterpath} {21,047}

\newcommand{\vulndissemcontractsinf} {17,278}
\newcommand{\vulndissemcontractsiter} {17,852}
\newcommand{\vulnsourcecontracts} {3,754}
\newcommand{\probdissemsnip} {616}
\newcommand{\probsourcesnip} {199}

\newcommand{\manualsamplesize} {100}
\newcommand{\manualvcv} {48}
\newcommand{\manualvcn} {3}
\newcommand{\manualncv} {1}
\newcommand{\manualncn} {22}

\newcommand{\manualnoclonesum} {26}
\newcommand{\manualvnv} {15}
\newcommand{\manualvnn} {3}
\newcommand{\manualnnv} {1}
\newcommand{\manualnnn} {7}

\newcommand{\snipreentrancy} {32}
\newcommand{\snipdos} {132}
\newcommand{\snipfrontrun} {159}
\newcommand{\sniptime} {96}
\newcommand{\snipshortaddress} {239}
\newcommand{\snipaccess} {21}
\newcommand{\snipover} {178}
\newcommand{\snipuncheckedcalls} {11}
\newcommand{\snipbadrng} {22}
\newcommand{\snipunknown} {1}

\newcommand{\conreentrancy} {240}
\newcommand{\condos} {3,524}
\newcommand{\confrontrun} {5,898}
\newcommand{\contime} {1,367}
\newcommand{\conshortaddress} {7,738}
\newcommand{\conaccess} {720}
\newcommand{\conover} {6,974}
\newcommand{\conuncheckedcalls} {11}
\newcommand{\conbadrng} {96}
\newcommand{\conunknown} {5}

\begin{abstract}

Ethereum smart contracts are executable programs deployed on a blockchain. Once deployed, they cannot be updated due to their inherent immutability. Moreover, they often manage valuable assets that are worth millions of dollars, making them attractive targets for attackers. The introduction of vulnerabilities in programs due to the reuse of vulnerable code posted on Q\&A websites such as Stack Overflow is not a new issue. However, little effort has been made to analyze the extent of this issue on deployed smart contracts.

In this paper, we conduct a study on the impact of vulnerable code reuse from Q\&A websites during the development of smart contracts and provide tools uniquely fit to detect vulnerable code patterns in complete and incomplete Smart Contract code. This paper proposes a pattern-based vulnerability detection tool that is able to analyze code snippets (i.e., incomplete code) as well as full smart contracts based on the concept of code property graphs. We also propose a methodology that leverages fuzzy hashing to quickly detect code clones of vulnerable snippets among deployed smart contracts. Our results show that our vulnerability search, as well as our code clone detection, are comparable to state-of-the-art while being applicable to code snippets. Our large-scale study on 18,660 code snippets reveals that \vulnsnipp~of them are vulnerable, out of which \probdissemsnip~can be found in \vulndissemcontractsiter~deployed smart contracts. These results highlight that the reuse of vulnerable code snippets is indeed an issue in currently deployed smart contracts.
\end{abstract}





\maketitle

\section{Introduction}

Smart Contracts are programs that enable complex computations in a distributed peer-to-peer network.
Ethereum is currently the most commonly used platform for deploying smart contracts on a public blockchain and manages its own digital currency called Ether. Ether is publicly tradeable and can be exchanged for contract executions on Ethereum's distributed computing engine, the quasi-Turing complete Ethereum Virtual Machine.
Unlike traditional programs, deployed smart contracts become immutable and thus cannot be changed.
This leads to a unique challenge in securing smart contracts. Since smart contracts cannot be updated after deployment, bug prevention has to be done during development.

With the rise of decentralized finance, smart contracts often handle assets that can easily be worth millions of USD.
Attackers keep exploiting vulnerabilities in smart contracts~\cite{rekt}.
The most infamous hack is the DAO hack that occurred in 2016. An attacker stole 3.6 million Ether~\cite{dao} by exploiting a reentrancy bug. This incident even led to a hard fork (a rewrite of Ethereum's transaction history).
Recent incidents have shown that long known vulnerabilities are still appearing in recent deployment contracts, e.g. the attack on the Hundred Fiance protocol on March 15, 2022 that led to 2,363 Ether being stolen (5.6 million USD~\cite{newreentrency} at the time). Vulnerable contracts continued to be attacked in 2023, e.g. LendHub losing around 6 million USD in 2023~\cite{lendhub}.

Since security-critical incidents on smart contracts keep reoccurring, recognizing and avoiding already known vulnerable coding practices is highly relevant.
As shown by previous research on Android applications~\cite{fischer2017stack}, code snippets from popular Q\&A websites can contain vulnerabilities but are often included by developers during development.
This form of software development is called ``Stack Overflow Driven Development'' (SODD). Developers search for code snippets on Q\&A websites such as Stack Overflow and copy code without understanding whether it is vulnerable.

This work studies whether or not this issue is prevalent in smart contract development. The toolchain we developed is necessary to enable the study, as current vulnerability detection tools either analyze smart contracts at the bytecode level or require compilable source code. Code snippets, however, can be incomplete contracts that are not compilable. This renders these tools unsuitable to study the reuse of code snippets.

In this paper, we present two tools: the Code property graph Contract Checker (\textsc{CCC}) and the Contract Clone Detector (\textsc{CCD}), which can handle code snippets as well as full smart contracts. We adapted an ANTLR grammar for Solidity to be able to parse code snippets and translate the resulting abstract syntax tree into a Code Property Graph (CPG) using a language-independent representation~\cite{weiss2022language}. While \textsc{CCC} loads the CPG into a graph database and runs a set of queries to find common smart contract vulnerabilities, \textsc{CCD} leverages N-gram matching and fuzzy hashing to quickly identify clones of vulnerable code snippets in a large dataset of deployed smart contracts.
We validate both tools against state-of-the-art smart contract vulnerability and code clone detection tools to ensure their viability for our study.
We combine both tools to identify vulnerable smart contract snippets on Stack Exchange~\cite{ethereum_stack_exchange} and Stack Overflow~\cite{stack_overflow}, and to check if any of the identified snippets are included in deployed smart contracts.
Our contributions are:

\begin{itemize}
    \item We propose a vulnerability detection tool that leverages code property graphs to detect common smart contract vulnerabilities in code snippets.
    \item We develop a clone detection tool that leverages fuzzy hashing to quickly detect similar code snippets across thousands of smart contracts.
    \item We perform the first systematic study that analyzes the impact of vulnerable code reuse from popular Q\&A websites, (e.g., Stack Overflow and Ethereum Stack Exchange), on Ethereum smart contracts\footnote{Tools and datasets are available for reproducibility at \href{https://github.com/Fraunhofer-AISEC/cpg-contract-checker}{github.com/Fraunhofer-AISEC/cpg-contract-checker} and \href{https://github.com/christoftorres/contract-clone-detector}{github.com/christoftorres/contract-clone-detector}}.
    \item We validate vulnerability patterns in \vulndissemcontractsiter~of \unqclonesofdissemsnip~deployed contracts that contain \probdissemsnip~vulnerable snippets from Q\&A websites, showing that vulnerable code reuse is an issue within the Ethereum ecosystem.
\end{itemize}

\section{Background}

\noindent
In this section, we provide background on Solidity, smart contract vulnerabilities, code property graphs, and code clones.

\subsection{Solidity and Smart Contracts}

\begin{lstlisting}[language=Solidity,caption={Example of a Solidity smart contract.},captionpos=b, ,label={lst:smallcontract}]
contract Parent {
  address owner;
  
  constructor() { owner = msg.sender; }
}

contract Main is Parent {
  uint state_var;
    
  constructor() { state_var = 0; }
    
  function() payable {...}
    
  function withdrawAll public onlyOwner() {
    msg.sender.call{value:this.balance}("");
  }
    
  modifier onlyOwner() {
    require(msg.sender == owner,"Not owner"); _;
  }
}
\end{lstlisting}

\noindent
Ethereum Virtual Machine (EVM)~\cite{wood2014ethereum} smart contracts are typically written in Solidity~\cite{solidity}, a high-level language with static typing, multiple inheritance, complex types, as well as nestable \texttt{\textcolor{blue}{contract}} constructs (e.g., comparable to \texttt{class} in OO programming languages). 
Since we analyze Solidity on source code level, we do not explain the translation of code into EVM bytecode.
\autoref{lst:smallcontract} highlights concepts used in Solidity to implement smart contracts. 
Constructors initialize contracts on deployment. 
In Line 4, the \texttt{owner} is set to the address of the Ethereum account that is deploying the smart contract (i.e., \texttt{\textcolor{violet}{msg}.\textcolor{violet}{sender}}). This is a common design pattern using Ethereum addresses as identity for access control. Line 10 declares a contract with the \texttt{\textcolor{blue}{constructor}} keyword, which is mandatory since Solidity 0.5 to avoid issues when misspelling function names. In Line 12, the contract declares a default function that is called when users invoke the contract without specifying a function name. The contents of the state variables, e.g. \texttt{state\_var}, are persisted after every transaction and represent the program state. The address in \texttt{owner} is used to control access to critical functionality. In Line 15, all funds of a contract are sent to the caller and the modifier \texttt{onlyOwner} declared in Line 18 is wrapped around some functions to ensure only the owner can execute them successfully. If the \texttt{\textcolor{blue}{require}} in Line 19 fails, the transaction fails, and all changes to involved contract are rolled back. 

\subsection{Smart Contract Vulnerabilities}\label{subsec:vulnbackground}

\noindent
The Ethereum blockchain is essentially a decentralized computer that stores information, executes functionality, and manages an internal currency (i.e., ether). The nature of a vulnerability, therefore, is a lack of integrity protection that allows an attacker to illicitly extract either ether from a smart contract, prevent the execution of contract functionality, or tamper with the information stored in a contract's program state. A number of such vulnerabilities have been presented in the past years \cite{atzei2017survey,perez2019smart}.
The Decentralized Application Security Project (DASP)~\cite{dasp} categorizes the 10 most common smart contract vulnerabilities: lacking restrictions to sensitive functionality~(\textbf{Access Control}); over- and underflows~(\textbf{Arithmetic}); use of predictable values for randomness~(\textbf{Bad Randomness}); operations that allow attackers to hinder contract execution~(\textbf{Denial of Service}); operations that can benefit a malicious user by preempting someone else's transactions~(\textbf{Front Running}); predictable code effects due to miners choosing the transactions' timestamp~(\textbf{Time Manipulation}); repeated/nested execution of a contract due to external contract calls~(\textbf{Reentrancy}); functions vulnerable to padding attacks in transaction addresses~(\textbf{Short Addresses}); unchecked return values of critical functions~(\textbf{Unchecked Low Level Calls}); and remaining vulnerabilities~(\textbf{Unknown Unknowns}).
Table~\ref{tab:daspeval} in our evaluation uses these categories to argue about the performance of our queries and limitations of the analysis.

\begin{figure*}[ht]
    \centering
    \includegraphics[width=\textwidth]{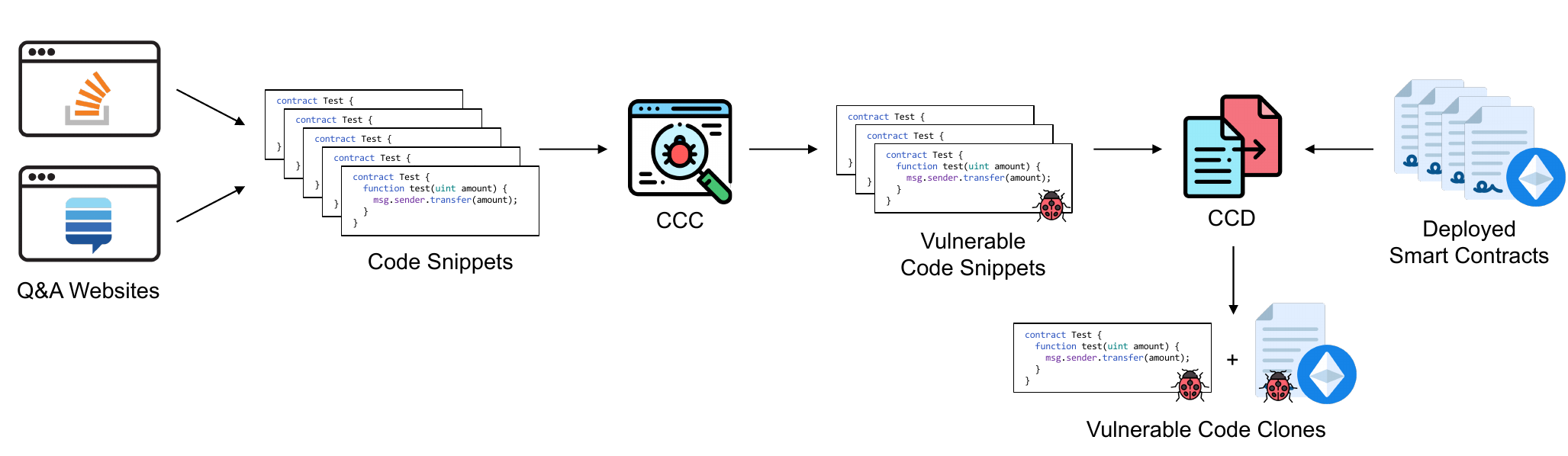}
    \caption{Processing pipeline of vulnerable code snippet identification (CCC) and code clone detection (CCD).}
    \label{fig:overview}
\end{figure*}

\subsection{Code Property Graphs}

\noindent
A Code Property Graph (CPG) is a directed attributed graph representing source code in the form of nodes and edges. Nodes embody syntactic elements of program code, whereas edges represent various semantics of a program. Nodes and edges have properties: key-value pairs that store additional information, such as code or location. The concept of a CPG was first introduced by Yamaguchi et al.~\cite{cpgfabian} and represents program syntax, control and data flow. Our work extends a CPG implementation~\cite{aisec_cpg} chosen for its flexible framework to implement language translation front-ends and semantics that are added by passes:
\vspace{-1mm}
\begin{itemize}
    \item \textbf{Syntax:} The program's Abstract Syntax Tree (\textbf{AST}) and its nodes form the basis for the graph's structure.
    \item \textbf{Order:} Evaluation Order Graph (\textbf{EOG}) edges are added to model control-flow and evaluation order. 
    \item \textbf{Data Flow:} Data Flow Graph (\textbf{DFG}) edges represent how data is transferred and processed in the program.
\end{itemize}

\begin{figure}
\centering
\includegraphics[width=0.9\columnwidth]{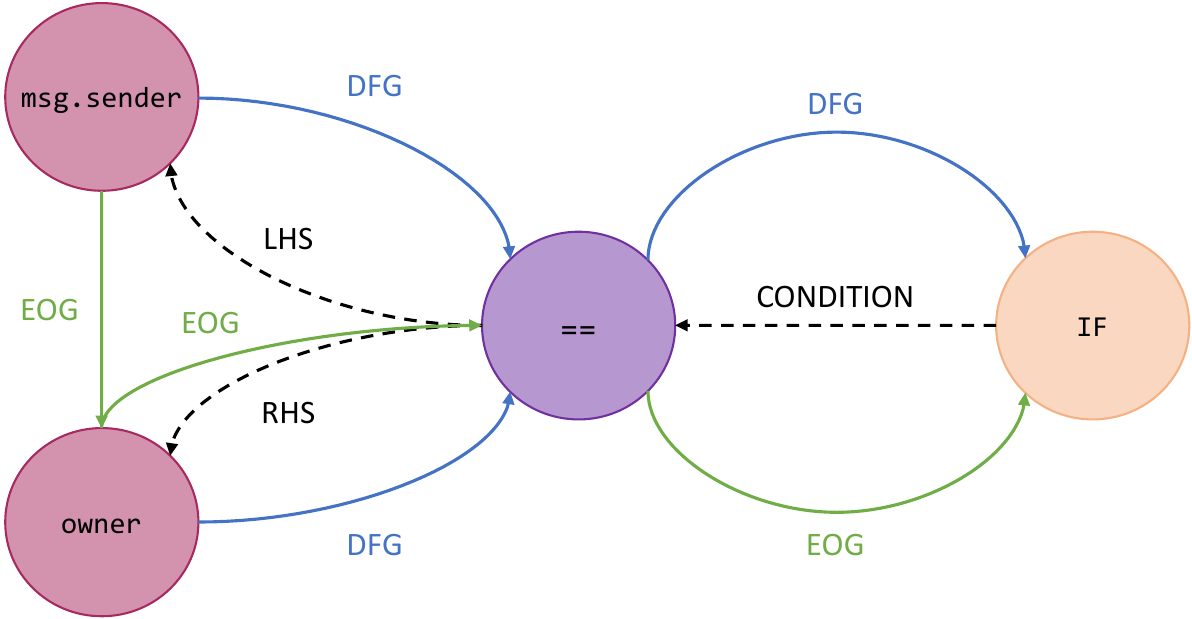}
\caption{Graph showing the programs syntax (dashed); the execution order EOG (green); and data flows DFG (blue).}
\label{fig:cpgexample}
\end{figure}

\noindent
Other edges are added for reference resolution, call targets, and the programs type system.
The resulting interconnected tree is persisted into a graph database and queried against specific vulnerability patterns. Figure~\ref{fig:cpgexample} shows an example graph when persisting \texttt{\textcolor{blue}{if}(\textcolor{violet}{msg}.\textcolor{violet}{sender} == owner)\{\}}. The EOG edges in green show that \textcolor{violet}{msg}.\textcolor{violet}{sender} is evaluated before the reference to the \texttt{owner}, and then are compared at \texttt{==} to evaluate the condition for the branching \texttt{IF}. The DFG edges in blue show how both references are used as input to evaluate \texttt{==} and are finally used to influence the branching \texttt{IF}. The remaining edges in gray show specific AST information. \texttt{LHS} shows the left-hand side and \texttt{RHS} shows the right-hand side of \texttt{==} that serves as \texttt{CONDITION} to the \texttt{IF}-statement.   

\subsection{Code Clones}

\noindent
Code clones are code fragments that are similar with respect to a clone type.
A code fragment is a contiguous segment of source code, with or without
comments. 
Fragments may contain any number of lines, statements, or functions.
To differentiate between different types of clones, we use the categorization of Roy and Cordy~\cite{roy2007survey}: 

\begin{itemize}
    \item \textbf{Type I (Exact Clone):} Identical fragments except for variations in comments and layout (e.g., whitespaces).
    \item \textbf{Type II (Renamed Clone):} Identical fragments except for variations in identifier names (e.g., variables) and literal values in addition to Type I differences.
    \item \textbf{Type III (Near-Miss Clone):} Syntactically similar fragments that differ at the statement level. The fragments differ in terms of added, modified, or removed statements, in addition to Type II differences.
    \item \textbf{Type IV (Semantic Clone):} Syntactically dissimilar fragments that implement the same functionality.
\end{itemize}

\noindent
Type I, II, and III clones indicate textual similarity, Type IV clones indicate semantic similarity.
Our focus is to detect code clones up to Type III, i.e. textual similarity, as these are related to SODD.
\section{Study Overview}

\noindent
The goal of this work is to identify whether Q\&A websites, e.g. Stack Overflow, contain vulnerable code snippets which are then used by software developers, then deploying vulnerable smart contracts.
In this section, we present our processing pipeline and describe the individual steps in more detail.
\noindent
As depicted in \figureautorefname{} \ref{fig:overview}, we start by extracting Solidity code snippets from Q\&A websites and identify vulnerable code snippets with our CPG Contract Checker (\textsc{CCC}), further discussed in Section \ref{sec:ccc}. Finally, we map vulnerable code snippets to deployed smart contracts using our Contract Clone Detector (\textsc{CCD}), presented in Section \ref{sec:ccd}. We developed both tools to resiliently analyze source code snippets, which other tools lacked.
\subsection{Limitations}\label{subsec:conceptuallimitaitons}
Our approach identifies code clones by comparing the similarity of snippets and deployed contracts. This method cannot prove that the snippet was copied from a Q\&A website. We reduce the likelihood of confusing causal directions in Section~\ref{subsec:experiment}. It is, however, still possible that a third source was responsible for the Q\&A snippet and the deployed contract. However, this limitation is shared with other works in the field and considered unlikely~\cite{fischer2017stack}. Proper operationalization to analyze causality can only be done by engaging in unethical dissemination of marked vulnerable code fragments and is, therefore, not done.
\section{Vulnerable Snippet Detection}
\label{sec:ccc}

\begin{figure}[t]
\centering
\includegraphics[width=\columnwidth]{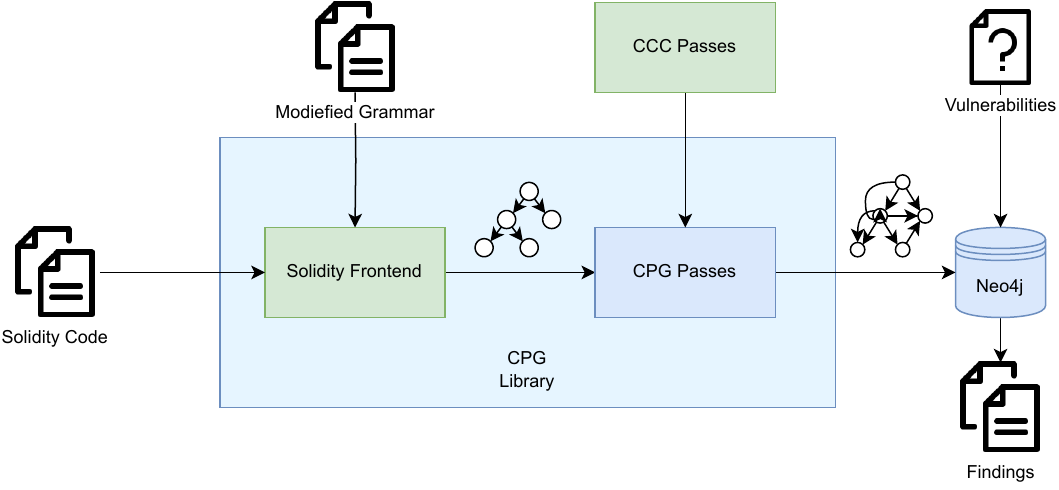}
\caption{Translation of Solidity code into a queriable graph and identification of vulnerabilities with CCC.}
\label{fig:architecture}
\end{figure}

\noindent
In this section, we describe the development of our vulnerability detection tool, the CPG Contract Checker \textsc{CCC}. We present our translation methodology of incomplete Solidity snippets into a CPG holding syntax, execution order, and data flows. Additionally, we will present the concept of our query design, how to express a vulnerability as query and discuss one query in detail. Figure~\ref{fig:architecture} shows our contributions in green and white, while the existing components are drawn in blue. The figure shows how we augment the CPG library~\cite{aisec_cpg} by implementing a Solidity Frontend to translate code into a CPG by using a modified grammar. We further enhance the program semantics added to the CPG-AST with our own set of \textsc{CCC} Passes. The graph is then persisted to a Neo4j graph database to identify vulnerabilities classified by DASP with our queries.

\subsection{Grammar Modifications}
\label{sec:grammar_modification}

\noindent
Language grammars represent a syntactically valid program that follows a program's structured hierarchy. Therefore, we have to modify an existing Solidity ANTLR grammar~\cite{antlr} to parse snippets:

\begin{itemize}
    \item \textbf{Unnesting of Hierarchy:} The Solidity language specifies a hierarchy at which program concepts, e.g. contracts, functions and statements, can appear. We lift several concepts to the global level which enables us to parse snippets at any level in the hierarchy.
    \item \textbf{Statement Termination:} We change the grammar to parse newline terminated statement to account for the mandatory ``\texttt{;}'' that was regularly missing in snippets.
    \item \textbf{Placeholders:} We ignore placeholders, e.g., ``...'', frequently added in snippets to signal additional code, to enable parsing as we already account for missing code in our queries.
\end{itemize}

\noindent
While these changes could lead to the incorrect parsing of valid Solidity code, we did not encounter such cases.

\subsection{Translation of Smart Contracts}
\noindent
We use the ANTLR parser with the modified grammar to get an AST which we translate into CPG nodes of the used library. To do so, we implement a language-specific translation component, the Solidity Language Frontend, supplementing the library. In this component, we translate a variety of expressions, statements, and declarations into their respective graph nodes. The declaration of \texttt{\textcolor{blue}{contract}} and state variables are translated into \texttt{Record} and \texttt{FieldDeclaration} as they are similar to the object-oriented concept of Classes and class-instance fields.
When a snipped is parsed, and the outer declarations of a \texttt{\textcolor{blue}{contract}} or function are missing, our frontend complements the translated AST with the inferred declarations.
In the following, we will present further peculiarities of the Solidity language and how we translated them into the CPG.

\subsubsection{Adding New Nodes}\label{subsec:newnodes}
Several concepts in the Solidity language could not be translated to preexisting CPG nodes. The execution of transaction code can be terminated and all changes to involved contracts reverted. To represent this program termination, we introduced a \texttt{Rollback} node and nodes for several Solidity expressions that can cause a rollback, e.g. \texttt{\textcolor{blue}{revert}} and \texttt{\textcolor{blue}{require}}. Further node types were added, such as nodes for emitting an event, i.e. persisting a message; special code blocks; an expression to specify gas limits and ether values, e.g., \texttt{contract.func\{\textcolor{blue}{value}: 1, \textcolor{blue}{gas}: 80\}()}.

\subsubsection{Modifier Expansion} 
Solidity allows to specify modifiers and adding them to function declarations. The purpose is to modularize validation of input data and access control. An example of a modifier can be seen in Line 18 of Listing~\ref{lst:smallcontract} and its use to modify  the behavior of a function in Line 14. When a modifier is used in a function header, the code of the function is wrapped in the modifier at every location of \texttt{\_;} in the modifier code. This allows to specify pre- and post-conditions.  
Modifier application cannot be modeled as a call to a function due to it allowing more complex wrapping of function code. Instead, we expand the AST of a function's body with the modifier code, creating copies of the modifier code for each of its applications. While recursive function inlining scales badly with call depth poses no issue for modifier expansion as developers use few modifiers per function, and modifiers cannot be nested. 

\subsubsection{Semantic extensions}
After we translate the parsed code into the CPG-AST, the passes of the CPG library add additional information based on language-independent heuristics, e.g., edges for data flows, evaluation order, and calls. To better represent the semantics of Solidity contracts, we supplement with our own passes:

\begin{itemize}
    \item Handlers are added to the DFG-Pass and EOG-Pass to model data flow and evaluation order for the newly created nodes mentioned in Section~\ref{subsec:newnodes}
    \item Additional DFG edges are added to the graph to cover indirect data flows necessary to find specific bugs.
\end{itemize}

\subsection{Query Design}\label{subsec:querydesign}

\noindent
We implement our search for vulnerable smart contract snippets as queries in the Cypher Query Language \cite{cypher}. With this declarative language, we can express graph patterns containing code syntax and semantics to find code that is considered problematic in its syntax or behavior. The following simplified query returns function parameters containing data that, at some point is written to a smart contract field, i.e., is persisted across transaction execution: \texttt{{\color{olive}MATCH} (p:Parameter)-[:DFG*]->(:Field) {\color{olive}RETURN} p}.
We conceptualized the following query components:

\begin{itemize}
    \item \textbf{Base patterns:} expressing information on syntax, data flow, execution order and types. This base pattern appears for every instance of the vulnerability, and one or more nodes are returned as vulnerability location.
    \item \textbf{Conditions of relevancy:} additional conditions or existential sub-queries of patterns that have to appear with the base pattern to qualify as vulnerability. One or more existential subqueries may be used disjunctively in the \texttt{\textcolor{blue}{WHERE}} clause of the query to express that it is sufficient for one subpattern to appear such that a vulnerability is found: \texttt{\textcolor{blue}{WHERE EXISITS} pattern A \textcolor{blue}{OR EXISTS} pattern B}
    \item \textbf{Mitigations and exceptions:} subqueries used to identify known exceptions to the base pattern or common mitigations of the searched vulnerability, e.g., locking, access control, or sanitization. Expressing these as a negated existential subquery reduces false positives: \texttt{\textcolor{blue}{WHERE NOT EXISITS} mitigation }
\end{itemize}

\subsection{Modeling Vulnerabilities}

\noindent
We implemented 17 queries in a best-effort approach to cover vulnerability patterns mappable to the 10 DASP categories mentioned in Section~\ref{subsec:vulnbackground}. Instead of learning vulnerable patterns from code done by previous work~\cite{fischer2017stack}, we write rule-based queries describing the underlying issues. 
In our query methodology, we build the above-mentioned base pattern from program patterns that can lead to the vulnerability described in the DASP category. We had to implement more than one query for a category if the base pattern could not be unified. We then considered what additional program behavior needs to be around the base pattern to constitute a potential implementation issue, e.g., external calls or sending of ether. Finally, we added patterns to exclude cases where the surrounding program behavior prevents the issue to the best of our knowledge. 
We will now show one of our queries and explain the query creation and the applied pattern concepts presented in Section~\ref{subsec:querydesign}.

\noindent
\newline
\textbf{Default Proxy Delegate.} A transaction targeting a function by name that does not appear in the contract is relayed over the default function to the delegate library. In 2017, the first Parity Wallet bug led to 153,037 ETH~\cite{paritymultisig} being stolen through an exploit of a delegate call in a default function. The snippet below shows the vulnerability where the delegation gave access to sensitive functions.

\vspace{0.2cm}
\begin{lstlisting}[language=Solidity,numbers=none,xleftmargin=0.7em,framexleftmargin=0.5em]
function() {lib.delegatecall(msg.data);}
\end{lstlisting}

\noindent
The query shown in Listing~\ref{lst:defaultProxyDelegateSimpl} identifies the vulnerability pattern. The first part constitutes the base pattern: A program path in a default function was found that does persist its results, i.e., does not end in a \texttt{Rollback} node and calls another contract through \texttt{\textcolor{blue}{delegatecall}}. The second part of the query forms the condition of relevancy: the caller can control the call target through the values in \texttt{\textcolor{blue}{msg.data}}. The final part shows how finding a mitigation technique leads to not producing a false positive: Along the path, there is a check on the content of message data that leads to an alternative path avoiding the call or leading to a rollback.

\begin{lstlisting}[language=Cypher,caption={Query finding call delegation vulnerabilities where inputs are not properly sanitized.},captionpos=b, ,label={lst:defaultProxyDelegateSimpl}, basicstyle=\scriptsize]
match p=(f:Function)-[:EOG*]->(c:Call)-[:EOG*]->(l)
where f.name=NULL and c.name in ['DELEGATECALL','CALLCODE']
and not exists ((l)-[:EOG]->()) and not 'Rollback' in labels(l)
-----------------------------------------------------
and (exists {({code:'msg.data'})<-[:ARGS]-(c)
} or exists{({code:'msg.data'})-[:DFG*]->()<-[:ARGS]-(c)})
-----------------------------------------------------
and not exists{
 df=(source {code: 'msg.data'})-[:DFG*]->(n)-[:EOG]->(ap) where n in nodes(p)
 and not exists {(of:Function|Call) where of in nodes(df)}
 and not exists ((source)<-[:BASE]-({code:'msg.data.length'}))
 and exists { d=(f)-[:EOG*]->(n)-[:EOG*]->(o) where not exists{ (op)-[:EOG]->() } and (not c in nodes(d) or 'Rollback' in labels(op))
}} return c
\end{lstlisting}
Note that Listing~\ref{lst:defaultProxyDelegateSimpl} is simplified for better readability. The full queries used in this work are listed in \appendixautorefname~\ref{sec:appendix}.

\begin{table*}
    \centering
    \caption{Comparison of CCC against other analysis tools using the SmartBugs Curated data set. Each vulnerability category is represented by a number of labeled vulnerabilities (\#). Results show the number of true positives and false positives for each tool and category as well as the totals. For the totals, precision and recall is provided.
    }
    \resizebox{\textwidth}{!}{%
    \begin{tabular}{l r | r r | r r | r r | r r | r r | r r | r r | r r | r r}
        \toprule
        & & 
          \multicolumn{2}{c|}{\textbf{\textsc{CCC}}} & 
          \multicolumn{2}{c|}{\textbf{\textsc{Confuzzius}}} & 
          \multicolumn{2}{c|}{\textbf{\textsc{Conkas}}} & 
          \multicolumn{2}{c|}{\textbf{\textsc{Mythril}}} & 
          \multicolumn{2}{c|}{\textbf{\textsc{Osiris}}} & 
          \multicolumn{2}{c|}{\textbf{\textsc{Oyente}}} & 
          \multicolumn{2}{c|}{\textbf{\textsc{Securify}}} & 
          \multicolumn{2}{c|}{\textbf{\textsc{Slither}}} & 
          \multicolumn{2}{c}{\textbf{\textsc{Smartcheck}}} \\
          \textbf{Vulnerability Category} & \textbf{\#} & TP & FP & TP & FP & TP & FP & TP & FP & TP & FP & TP & FP & TP & FP & TP & FP & TP & FP \\
        \midrule
        Access Control      & 21 & 
          10 & 2 & 2 & 7 & 0 & 0 & 7 & 4 & 0 & 0 & 0 & 0 & 0 & 1 & 6 & 1 & 2 & 0 \\
        Arithmetic          & 23 & 
          17 & 1 & 15 & 1 & 18 & 3 & 15 & 2 & 19 & 2 & 14 & 4 & 0 & 0 & 0 & 0 & 0 & 0 \\
        Bad Randomness      & 31 & 
          12 & 2 & 2 & 12 & 0 & 0 & 0 & 4 & 0 & 0 & 0 & 0 & 0 & 0 & 0 & 0 & 0 & 0 \\
        Denial of Service   &  7 &  
           6 & 1 & 0 & 0 & 0 & 0 & 1 & 0 & 0 & 6 & 0 & 1 & 0 & 0 & 1 & 0 & 0 & 0 \\
        Front Running       &  7 & 
           2 & 1 & 1 & 1 & 2 & 0 & 0 & 0 & 2 & 2 & 2 & 2 & 2 & 4 & 0 & 0 & 0 & 0 \\
        Reentrancy          & 32 & 
          28 & 3 & 28 & 11 & 28 & 44 & 26 & 1 & 28 & 20 & 28 & 0 & 28 & 7 & 0 & 8 & 0 & 0 \\
        Short Addresses     &  1 &  
           1 & 1 & 0 & 0 & 0 & 0 & 0 & 0 & 0 & 0 & 0 & 0 & 0 & 0 & 0 & 0 & 0 & 0 \\
        Time Manipulation   &  7 &  
           7 & 2 & 0 & 0 & 5 & 5 & 2 & 2 & 1 & 1 & 0 & 0 & 0 & 0 & 2 & 1 & 1 & 1 \\
        Unchecked Low Level Calls & 75 & 
          75 & 0 & 52 & 1 & 62 & 0 & 46 & 4 & 0 & 0 & 0 & 0 & 67 & 14 & 51 & 9 & 61 & 1 \\
        \midrule
        \textbf{Total} & \textbf{204} & 
          158 & 13 & 100 & 33 & 115 & 52 & 97 & 17 & 50 & 31 & 44 & 7 & 97 & 26 & 60 & 19 & 64 & 2 \\
        \midrule
        \textbf{Precision/Recall} & \textbf{} & 
          92.3\% & 77.4\% & 75.2\% & 49.0\% & 68.9\% & 56.4\% & 85.1\% & 47.5\% & 61.7\% & 24.5\% & 86.3\% & 21.6\% & 78.9\% & 47.5\% & 75.9\% & 29.4\% & 97.0\% & 31.4\% \\
        \bottomrule
    \end{tabular}
    }
    \label{tab:smartbugseval}
\end{table*}

\begin{table}[t]
    \caption{Performance Comparison of CCC against SmartBugs Curated and extracted vulnerable snippets (\textit{Functions} \& \textit{Statements}; cf. \ref{sec:ccc-exp-setup}). 
    }
    \label{tab:ccc-snippet-performance}
    \centering
    \resizebox{\columnwidth}{!}{%
        \begin{tabular}{l | r r | r r | r r}
            \toprule
             & \multicolumn{2}{c|}{\textbf{Original}} & \multicolumn{2}{c|}{\textbf{Functions}} & \multicolumn{2}{c}{\textbf{Statements}} \\
             & TP & FP & TP & FP & TP & FP \\
            \midrule
            \textbf{Total} (204) & 
                158 & 13 & 146 & 6 & 121 & 2 \\
            \midrule
            \textbf{Precision/Recall} & 
                92.3\% & 77.4\% & 96.0\% & 71.5\% & 98.3\% & 59.3\%  \\
            \bottomrule
        \end{tabular}
    }
\end{table}

\subsection{Limitations}

\noindent
Our vulnerability detection is limited to Solidity source code as we try to analyze snippets that are generally not compilable. Only 3.6\% of snippets contain assembly, which we therefore did not model. Our approach is specifically developed to handle incomplete code, which needs analysis on the source code level. While pattern-based analysis on CPGs can leverage data flow and control flow information in addition to syntax, the results are more prone to false positives than specialized detection tools using symbolic execution and SAT-solving. However, more sophisticated tools are not applicable to snippet analysis.

\subsection{Evaluating CCC}
\label{subsec:smartbugs}

\noindent
We evaluate the performance of CCC against other well-established smart contract analysis tools using a dataset of labeled smart contract vulnerabilities.
Moreover, we assess CCC's ability to detect vulnerabilities within Solidity code snippets.

\subsubsection{Experimental Setup}
\label{sec:ccc-exp-setup}
We leveraged SmartBugs 2.0.9 \cite{diAngelo-2023-SmartBugs2} to compare CCC against other state-of-the-art analysis tools. 
SmartBugs is a framework to compare the analysis of Ethereum smart contracts.
It currently supports 20 tools for analyzing Solidity source code and/or EVM bytecode.
We integrated CCC as an additional analysis tool into SmartBugs to automate the analysis and comparison to other tools.
With our focus on Solidity source code we have selected Confuzzius \cite{torres2021confuzzius}, Conkas \cite{conkas}, Mythril \cite{mythril}, Osiris \cite{torres2018osiris}, Oyente \cite{oyente}, Securify \cite{securify}, Slither \cite{Slither} and SmartCheck \cite{tikhomirov2018smartcheck} as analysis tools for the performance evaluation. 

We assess the performance of CCC and other analysis tools using the SmartBugs Curated data set \cite{durieux2020empirical, smartbugs-curated}. 
SmartBugs Curated consists of $143$ Solidity files divided into $10$ vulnerability categories based on the DASP taxononmy \cite{dasp}. 
Vulnerabilities are labeled accordingly for each category within each Solidity file. 
We excluded the category "Other" from SmartBugs Curated consisting of three Solidity files because it is associated with the vulnerability category of "Unknown Unknows" within the DASP taxonomy. 
This DASP category of "Unkown Unkowns" is a fallback category for any tool finding that does not fit any of the other nine DASP categories. 
Thus, an assessment of "Other" would likely be skewed because any tool finding unrelated to the labeled vulnerabilities in "Other" would falsely count towards the number of false positives. 
Therefore, we decided to exclude the category "Other" and the few labeled vulnerabilities of the category. 
Our final dataset consists of $140$ Solidity files across $9$ vulnerability categories with a total of $204$ labeled vulnerabilities.

We also modified the SmartBugs Curated datset to assess CCC's ability to handle code snippets. 
We derived two additional datasets called \textit{Functions} and \textit{Statements}. 
For the dataset \textit{Functions} we extracted functions containing labeled vulnerabilities and stored each function into its own file. 
For the dataset \textit{Statements} we acted likewise and extracted labeled statements with up to a total of five statements around it, if sufficiently many statements were available. 
The extracted statements did not include function headers. 
The resulting datasets contain non-compilable code snippets with potential vulnerabilities. 
Both datasets have the same number of labels as the original SmartBugs Curated dataset.

\subsubsection{Comparative Results}
\label{sec:ccc-exp-results}
\autoref{tab:smartbugseval} presents our results on detecting labeled vulnerabilities within the SmartBugs Curated dataset.
The results show the number of true positives and false positives as well as the resulting precision and recall. Note that the SmartBugs test sets of a category only contains labels for that category. In several cases our and other tools correctly reported issues in the test set of a different category.
We only counted findings as false positives if they were reported in the matching test set, assuming that vulnerabilities of the test sets category are correctly labeled. This restriction on false positive counting was done for all tools and allows to compare false positive numbers.

CCC identifies 158 of the labeled vulnerabilities but also reports 13 vulnerabilities of a category in the matching test set; as their locations are not labeled, we counted them as false positives. Our qualitative analysis of these false positives showed that we incorrectly flagged two cases where we did not recognize complex access control protecting sensitive functionality. We mislabeled two uses of a block number as bad RNG computation, where it was used legitimately and not to compute a random number. We mislabeled one expensive loop that was not controllable by an attacker. We had one duplicate report of another false positive and found one vulnerability that SmartBugs did not label as its exploitation is unlikely. The remaining six cases of false positives were when our tool and the SmartBugs labeling mismatched the location of the same vulnerability. This happens when calls and modifiers are involved, and every report location along the call chain is legitimate. In other cases, we reported the problematic operation while the condition or function not preventing access to it was labeled. However, other tools share this issue of mismatching the finding location with the SmartBugs label, although we cannot make claims on the frequency of this issue.

The comparison with other analysis tools across the nine vulnerability categories in \autoref{tab:smartbugseval} show that  CCC performs best in reporting labeled vulnerabilities with $158$ out of $204$ resulting in a recall of $77.4\%$. 
These reported findings are by far the highest among the considered analysis tools. 
The second best tool in identifying vulnerabilities is Conkas with $115$ correctly identified vulnerabilities. 
However, Conkas has considerably more false positives with $52$ compared to CCC's $13$, which is also reflected in the lower recall of $56.4\%$. 
The good performance of CCC is also highlighted by the second highest precision of $92.3\%$. 
This precision is only second to SmartCheck with a precision of $97.0\%$. 
However, SmartCheck reports only $64$ true positive out of $204$ labeled vulnerabilities, which is considerably smaller than CCC's $158$. 
In more detail, CCC out-performs the other analysis tools in 6 out of nine vulnerability categories with respect to found true positives. 
In the categories "Front Running" and "Reentrancy", CCC is on par with the other analysis tools. 
Finally, in the category "Arithmetic", CCC places third after Osiris and Conkas. 
Moreover, CCC reports findings across all nine vulnerability categories, which is unique among the evaluated tools. 
Other tools cover at most six categories. 
In summary, CCC performance is comparable to current state-of-the-art analysis tools for Ethereum smart contracts. 
It correctly identifies most vulnerabilities with a precision and a recall among the highest of all evaluated tools. 

\subsubsection{Results on Derived Datasets}
\label{sec:ccc-exp-results} 
\autoref{tab:ccc-snippet-performance} shows the results of the second part of the evaluation of CCC. Here we have derived the two datasets \textit{Functions} and \textit{Statements} from SmartBugs Curated (cf. \ref{sec:ccc-exp-setup}) to reflect vulnerable code snippets. 
Both datasets represent Solidity code snippets, which we use to assess CCC's performance to detect vulnerabilities in code snippets such as those collected from Q\&A websites. 
The results are presented in \autoref{tab:ccc-snippet-performance}.
For the \textit{Functions} dataset CCC detects almost the same number of vulnerabilities ($146$) compared to the original SmartBugs Curated dataset ($158$). 
At the same time CCC finds fewer false positives, but more false negatives. 
These numbers result in an overall higher precision but lower recall. 
The results are likely due to simpler vulnerability patterns when analyzing only individual functions. 
However, vulnerabilities from interactions between functions are missed. 
For the \textit{Statements} dataset, the number of detected vulnerabilities decreases to $122$. 
Precision increases to $98.3\%$, while recall decreases to $59.3\%$. 
These results are likely due to missing context information that CCC uses to judge if potentially vulnerable statements can lead to exploitable behavior. 
Overall, our results show that CCC can identify vulnerabilities in incomplete and non-compilable code such as Solidity code snippets. 
The detection performance decreases with code snippets due to missing context information that would otherwise support the vulnerability analysis. 
At the same time, CCC is able to analyze code snippets, whereas other analysis tools require compilable code.

\section{Code Clone Detection}
\label{sec:ccd}

\begin{figure*}
    \centering
    \includegraphics[width=\textwidth]{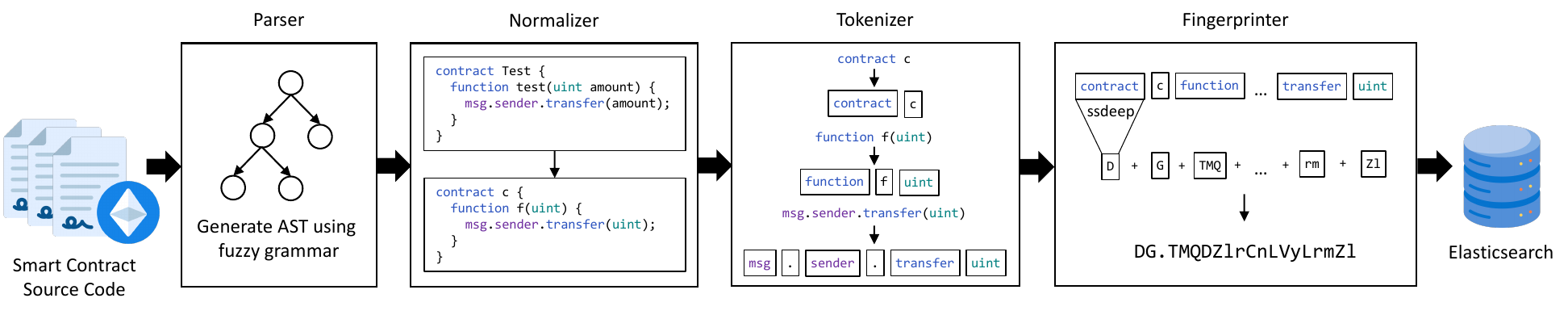}
    \caption{Pipeline architecture of our code clone detection.}
    \label{fig:morpheus_architecture}
\end{figure*}

\noindent
In this section, we explain our approach towards detecting code clones of code snippets in Solidity smart contracts. \figureautorefname{} \ref{fig:morpheus_architecture} provides an overview of the overall pipeline architecture of our Contract Clone Detector (\textsc{CCD}).

\subsection{Parsing}

\noindent
As a first step, we need to parse the smart contract source code. To deal with code clones of Type I, we remove all whitespaces, new lines, and comments from the Solidity source code. 
Afterwards, we leverage our own parser that uses a custom Solidity ANTLR grammar that is capable of parsing incomplete Solidity source code (i.e., code snippets, see Section \ref{sec:grammar_modification}). The parser generates an AST, which is essentially an XML parse tree that provides a structural representation of the smart contract.

\subsection{Normalization}

\noindent
After generating the AST we normalize the source code by traversing the AST and renaming any identifiers such as contract declarations, variable names, parameters, etc.
All contract declarations are renamed to ``\texttt{c}'', all library names are renamed to ``\texttt{l}'', all function names are renamed to ``\texttt{f}'', and all modifiers are renamed to ``\texttt{m}''. All parameters and variables are renamed according to their declared type. For code snippets with missing type declarations we simply use the default type \texttt{uint}. For example, the following code snippet:

\vspace{0.2cm}
\begin{lstlisting}[language=Solidity,numbers=none,xleftmargin=1.0em,framexleftmargin=0.5em]
contract Test {
  function test(uint amount) {
    msg.sender.transfer(amount);
  }
}
\end{lstlisting}
\vspace{0.2cm}
\noindent
is normalized to the following code:
\vspace{0.2cm}
\begin{lstlisting}[language=Solidity,numbers=none,xleftmargin=1.0em,framexleftmargin=0.5em]
contract c {
  function f(uint) {
    msg.sender.transfer(uint);
  }
}
\end{lstlisting}
\vspace{0.2cm}

\noindent
Moreover, we replace string literals with the keyword \texttt{string
Literal}. Numeric constants are left untouched since a difference in these may have an impact on whether the smart contract is vulnerable or not, thus normalizing numeric constants could yield false positives when matching. Finally, we also normalize function visibility, such as \texttt{public} or \texttt{view}, by simply removing them. 
The renaming of identifiers, string literals, and function visibility allows us to deal with code clones of Type II, where developers copied some code blocks but modified the names of the variables or functions.

\subsection{Tokenization}

\noindent
In this step, we split the normalized source code into tokens. We ignore state variable declarations as well as event declarations and only focus on contract declarations and function declarations as well as function-level statements. 
Afterwards, we simply divide the code based on symbols (e.g., ``+'', ``-'', ``;'', or brackets). This allows us to only preserve the context that is relevant to us. For example, the following statement: \texttt{\textcolor{violet}{msg}.\textcolor{violet}{sender}.\textcolor{blue}{transfer}(\textcolor{teal}{uint})} is tokenized into the following six tokens: `\texttt{\textcolor{violet}{msg}}', `\texttt{.}', `\texttt{\textcolor{violet}{sender}}', `\texttt{.}', `\texttt{\textcolor{blue}{transfer}}', `\texttt{\textcolor{teal}{uint}}'. These tokens are then used to generate a unique fingerprint.

\subsection{Fingerprint Generation}
\label{sec:fignerprint_generation}

\begin{figure}
    \centering
    \includegraphics[width=\columnwidth]{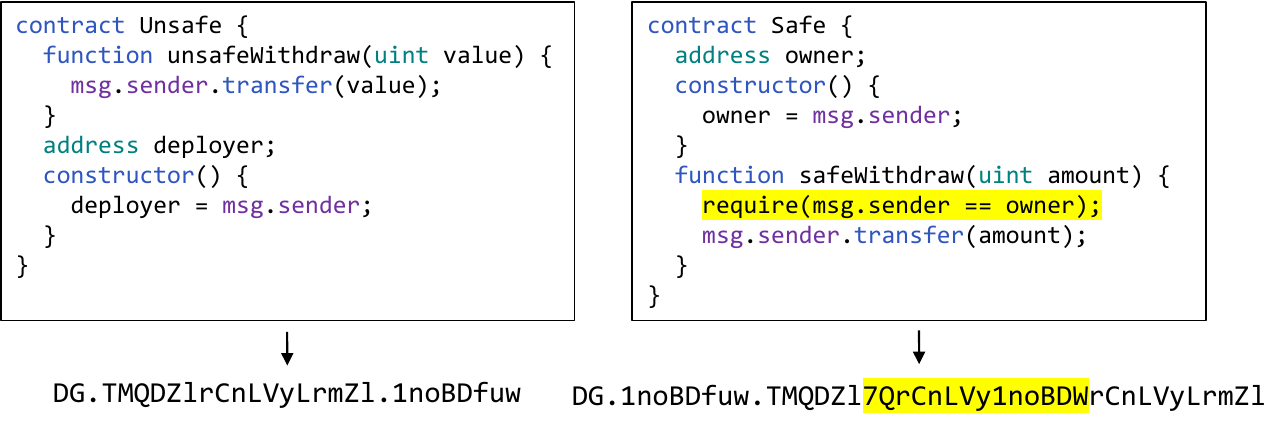}
    \caption{Two similar snippets and their fingerprints.}
    \label{fig:fingerprint_example}
\end{figure}

\noindent
We leverage fuzzy hashing \cite{Kornblum06} to condense the normalized and tokenized source code to a so-called fingerprint, a much shorter representation of the original source code. We then calculate the edit distance between two fingerprints to measure their similarity and to identify if they are code clones or not. 
Unlike traditional hash functions, fuzzy hashing first splits a sequence into smaller pieces and uses
a piece-wise hash function to compute a hash for each piece. The final fingerprint is generated by concatenating the piece-wise hashes together.
As opposed to a traditional hash function where one little change in the input results in an entirely different hash, in the case of fuzzy hashing, only the modified pieces will result in
different piece-wise hashes, hence, the resulting fingerprint will not be entirely different but will still look very similar. \figureautorefname{} \ref{fig:fingerprint_example} provides a concrete example of two similar code snippets. We observe that the newly added line (highlighted in yellow) only modifies a part of the fingerprint and that the rest remains the same.
Fuzzy hashing reduces memory usage and computation time as the matched sequences are much smaller.

One of the main challenges of fuzzy hashing is to determine the optimal boundary of each piece. One of the most widely used fuzzy hashing tools is \texttt{ssdeep} \cite{JAKOBS2022301402}, which splits by default any sequence of bytes into chunks (i.e., pieces) of seven bytes. 
We leverage \texttt{ssdeep} to transform our tokenized source code into a fingerprint. However, instead of concatenating all tokens together and feeding them directly as one sequence to \texttt{ssdeep}, we feed each token one by one to \texttt{ssdeep} and concatenate the resulting fuzzy hash to a final fingerprint. This allows us to enforce context on the fuzzy hashes and hence the final fingerprint. The final fingerprint is a sequence of base-64
characters, where function code is separated from one another using a period (i.e., ``.'') and contracts are separated using a colon (i.e., ``:''). This is useful later when matching fingerprints of functions and contracts, irrespective of their order in the code (see Section \ref{sec:fingerprint_matching}). 

\subsection{Fingerprint Matching}
\label{sec:fingerprint_matching}

\noindent
Finally, we aim to match a code snippet against thousands of other snippets via their fingerprint. We could simply compute the edit distance between every fingerprint and return those code snippet pairs whose fingerprints have a small edit distance (i.e., are very similar). However, using this approach, we would face two challenges: 
\begin{itemize}
    \item \textbf{Execution Time.} Computing the edit distance for every pair of fingerprints is very expensive in terms of computation time and would render the matching inherently slow;
    \item \textbf{Code Order.} If two code fragments contain the exact same function and contract code, but their order is swapped, then this will result in two distinct fingerprints and a simple edit distance calculation will result in a low similarity score.
\end{itemize}
We solve the first challenge by splitting up the fingerprint into N-grams of size $N$ and storing them into an Elasticsearch database. When matching a particular fingerprint, we first query the Elasticsearch database and retrieve only those fingerprints which include the same $N$-grams as the matching fingerprint up to a given threshold $\eta$. For example, an $N$-gram threshold of $\eta = 0.5$ means that only fingerprints that include at least 50\% of the same $N$-grams as the fingerprint that is being searched for will be retrieved. This step is very fast as Elasticsearch already automatically splits strings into $N$-grams and indexes those for fast retrieval. This allows us to apply some filtering and therefore reduce the number of candidates on which we have to compute the edit distance. 

To solve the second challenge, we do not directly compute the edit distance on two given fingerprints. We split each fingerprint into several sub-fingerprints. As described in Section \ref{sec:fignerprint_generation}, our fingerprints contain non-base-64 characters such as ``.'' and ``:'' in order to separate function implementations and contract definitions. The idea is to individually match fingerprints of function implementations irrespective of their order across code snippets. 
We compute the similarity score for two sub-fingerprints $s_1$ and $s_2$ as follows:

$$
\delta(s_1, s_2) = \frac{max(len(s_1), len(s_2)) - d(s_1, s_2)}{max(len(s_1), len(s_2))} * 100,
$$

\noindent
\\
where $d$ is the edit distance between two strings.
The final similarity score $\epsilon$ is computed as defined in Algorithm \ref{alg:similarity_score}.
The idea is to match every sub-fingerprint from fingerprint $f_1$ with all the sub-fingerprints from fingerprint $f_2$ and to take the average of the sum of the highest similarity scores $\delta$ for every sub-fingerprint from $f_1$.

\begin{algorithm}
\caption{Order-Independent Similarity Score}
\label{alg:similarity_score}
\small
\begin{algorithmic}
\State \textbf{Input:} $f_1, f_2$; \textbf{Output:} $\epsilon$
\State $i \gets \{\}$
\ForAll{$s_1 \in f_1$}
\State $j \gets \{\}$
\ForAll{$s_2 \in f_2$}
\State $j \gets j + \delta(s_1, s_2)$
\EndFor
\State $i \gets i + max(j)$
\EndFor
\State $\epsilon \gets \frac{sum(i)}{len(i)}$
\end{algorithmic}
\end{algorithm}

\subsection{Limitations}

\noindent
Our approach is not capable of detecting semantic code clones (i.e., clones of Type IV). However, it is able to detect syntactically similar code clones (i.e., clone of Type I, II, and III.). We argue that this is sufficient for the scope of this paper as we aim to detect code snippets which have been copied and pasted by developers and where we assume they might have added comments, changed variable names, or even slightly modified the code snippets by adding or removing some lines of code from the original snippet. However, we do not assume that developers created syntactically entirely different code from code on Q\&A websites.

\subsection{Evaluating CCD}

\noindent
We evaluate the performance of CCD by comparing it to \textsc{SmartEmbed} \cite{gao2019smartembed}, a state-of-the-art clone detection tool for smart contracts that leverages structural code embeddings. Please note that, as opposed to CCD, \textsc{SmartEmbed} requires complete code that compiles and cannot analyze incomplete code snippets out-of-the-box. 

\subsubsection{Experimental Setup}

We compare both tools using a labeled dataset published by Torres et al.~\cite{TorresSS19,CaminoTBS20} that contains smart contract honeypots. These are scams deployed by malicious parties that try to trick users into sending their funds to the honeypot smart contract, keeping the funds and not returning back the hoped return on investment to the users. Torres et al. identified nine different types of such honeypots. Honeypots are a good fit to compare code clones as their code is often very similar to each other. This is because honeypot creators keep reusing the same ``technique'' and only slightly modify the code around it to not be entirely identical to a previous contract. The dataset provided by Torres et al. contains in total $379$ honeypot smart contracts. 

In our experiment, we computed for each smart contract contained in the dataset its similarity score to all other smart contracts contained in the dataset using both our tool and \textsc{SmartEmbed}. 
For \textsc{SmartEmbed}, we used a similarity score of $0.9$ (as recommended by the authors) to decide whether a smart contract is a clone or not. For our own clone detection tool, we tried different combinations of N-gram sizes, N-gram thresholds, and similarity thresholds (see \tableautorefname{} \ref{tab:parameters} in Appendix \ref{sec:appendix_code_clone_detection_parameters}).
The best combination of both precision and recall was achieved using an N-gram size of 3, an $\eta$ threshold of $0.5$, and an $\epsilon$ threshold of $0.7$ (see \figureautorefname{} \ref{fig:parameter_results} in Appendix \ref{sec:appendix_code_clone_detection_parameter_results}).

\subsubsection{Results}

\begin{table}
    \caption{True/False positive comparison between \textsc{SmartEmbed} and \toolname{}.}
    \label{tab:tp_fp_comparison}
    \centering
    \small
    \begin{tabular}{l r r | r r}
        \toprule
        & \multicolumn{2}{c|}{\textbf{\textsc{SmartEmbed}}} & \multicolumn{2}{c}{\textbf{\toolname{}}} \\
        \textbf{Honeypot Type} & \textbf{TP} & \textbf{FP} & \textbf{TP} & \textbf{FP} \\
        \midrule
        Balance Disorder             & 121 & 29  & 210 & 104 \\
        Type Deduction Overflow      &  21  & 0   &  9  & 0 \\
        Hidden Transfer              & 166 & 108  & 136  & 0 \\
        Unexecuted Call              &  14  & 6  &  14  & 2 \\
        Uninitialised Struct         & 410  & 2  & 397  & 2 \\
        Hidden State Update          & 6,804 & 156 & 6,912 & 139 \\
        Inheritance Disorder         & 397 & 90  & 415 & 49 \\
        Skip Empty String Literal    &  27 & 19  &  27 &  0 \\
        Straw Man Contract           & 443 &  8  & 616  & 6 \\
        \midrule
        \textbf{Total} & \textbf{8,403} & \textbf{418} & \textbf{8,736} & \textbf{302} \\
        \bottomrule
    \end{tabular}
    \vspace*{-0.7\baselineskip}
\end{table}

\tableautorefname{} \ref{tab:tp_fp_comparison}, lists the number of true positives (TP) and false positives (FP) for both, \textsc{SmartEmbed} and our tool \toolname{} across different types of honeypots. Overall, our tool reports less false positives (i.e., $302$) as compared to \textsc{SmartEmbed} (i.e., $418$), and more true positives than \textsc{SmartEmbed} (i.e., $8,736$ as opposed to $8,403$). For honeypots of type balance disorder, our tool reports more false positives than \textsc{SmartEmbed}, on the other hand our tool detects more true positives than \textsc{SmartEmbed}. For honeypots of type type deduction overflow, hidden transfer, and uninitialized struct, our tool reports less true positives, but in the case of hidden transfer it reports zero false positives as opposed to $108$ false positives reported by \textsc{SmartEmbed}.
Overall, our tool achieves a higher precision (0.9666 vs. 0.9526), recall (0.2563 vs. 0.2465), and F1-score (0.4052 vs. 0.3917) than \textsc{SmartEmbed}.
\section{Vulnerable Code Reuse}
\noindent
In this section, we combine our vulnerability detection and code clone detection to study vulnerable code reuse in smart contracts that originate from Q\&A websites.

\subsection{Data Collection}

\begin{table}
    \caption{Overview of our Solidity code snippet dataset.}
    \label{tab:snippet_dataset}
    \centering
    \begin{adjustbox}{width=\columnwidth,center}
    \begin{tabular}{l r|r r r | r}
        \toprule
        \textbf{Q\&A Website} & \textbf{Posts} & \textbf{Snippets} & \textbf{Solidity} & \textbf{Parsable} & \textbf{Unique} \\
        \midrule
        Stack Overflow & 7,370 & 12,111 & 7,116 & 5,305 & 5,187\\
        Ethereum Stack Exchange & 18,283 & 27,323 & 18,609 & 14,565 & 13,473 \\
        \midrule
        \textbf{Total} & \textbf{25,653} & \textbf{39,434} & \textbf{25,725} & \textbf{19,870} & \textbf{\analyzedsnippets} \\
        \bottomrule
    \end{tabular}
    \end{adjustbox}
\end{table}

We gather Solidity code snippets posted on developer Q\&A websites by crawling Stack Overflow \cite{stack_overflow} and the Ethereum Stack Exchange \cite{ethereum_stack_exchange} for posts created until June 30, 2023 with the tag ``solidity''. Overall, we collect 39,434 code snippets across 25,653 posts, where 12,111 snippets originate from 7,370 posts on Stack Overflow and 27,323 snippets originate from 18,283 posts on the Ethereum Stack Exchange (see \tableautorefname{} \ref{tab:snippet_dataset}). However, not all snippets are necessarily Solidity code snippets. Some contain JavaScript code or some form of pseudo-code that has been tagged as ``solidity'' by the author. 

We filter out snippets that are not related to Solidity by checking if they include unique Solidity keywords. JavaScript has $124$ keywords, whereas Solidity has $251$ keywords. However, both languages have several keywords in common such as ``\texttt{var}'' or ``\texttt{public}''.
After removing common keywords, we are left with $166$ unique Solidity keywords.
After filtering out snippets that do not include at least one of the $166$ keywords, we are left with {25,725} snippets, {7,116} from Stack Overflow and {18,609} from Ethereum Stack Exchange.

Despite using a modified grammar that allows to parse incomplete code, some snippets simply cannot be parsed, because they include text that is not valid Solidity code (e.g., a mix of pseudocode with valid Solidity keywords). Hence, we filter out all snippets, which cannot be parsed using our grammar. This results in {19,870} parsable snippets, where {5,305} are from Stack Overflow and {14,565} are from the Ethereum Stack Exchange. Our grammar manages to parse {3,133} more snippets than the standard Solidity grammar. 
Majority of the parsed snippets contain contract definitions ({54,2\%}), followed by snippets with only function definitions ({38\%}), and finally snippets that contain only statements ({7.8\%}).
Moreover, the maximum number of lines of code of the parsed snippets is {775}, whereas the mean is {22}, the median {12}, and the minimum is {1}.
Lastly, we removed all duplicates and were left with a total of \analyzedsnippets~unique snippets that can be used for our study, of which {5,187} are from Stack Overflow and {13,473} are from the Ethereum Stack Exchange. 

Our final goal is to map vulnerable snippets to deployed smart contracts. We leverage the Ethereum Smart Contract Sanctuary \cite{smart_contract_sanctuary} dataset, which contains the source code of deployed smart contracts that are verified on Etherscan. 
The dataset consists of {323,328} smart contracts with Solidity source code deployed until {July 14, 2023}.
Most of the contracts ({59\%}) were deployed using v0.8 of the Solidity compiler, {16\%} using v0.6, {13\%} using v0.4, {7.4\%} using v0.5, and roughly {4\%} using v0.7. Between May and June of 2023, majority ({91\%}) were deployed using v0.8 (latest version), while {9\%} were deployed using an older version, which makes the latter susceptible to vulnerabilities (e.g., integer overflows \cite{solidity_0.8}).

\subsection{Popularity and Appearance Frequency}
\label{subsec:correlation}
In this section, we analyze whether a snippet in a popular Q\&A post has an increased likelihood of appearing in a real-world smart contract. We define the popularity as the number of views \textit{v} on a Q\&A post. If an effect exists that leads to developers adopting snippets from Q\&A websites, we expect to see a correlation between the number of views \textit{v} and the number of contracts containing the snippet. If no correlation is measured, either no effect can be assumed, or a different variable determines whether a snippet is copied into a contract.

We want to differentiate between snippets that are more or less likely to be postings of already deployed contracts or third-party sources. Snippets that are more likely to be postings of other sources should have a weaker correlation between \textit{v} and \textit{nr} than those that are less likely and, therefore, are more likely to represent cases of copy\&paste code from the snippet to contract. Consequently, we expect higher correlation as the temporal restrictions between snippets-contract pairings increase in the following sets:

\begin{itemize}
\item \emph{All Snippets:} A snippet and all contracts containing it are considered, those posted before and after the snippet's appearance on a Q\&A website.
\item \emph{Disseminator:} Snippets for which we identified at least one contract that was deployed after the snippet was posted. We only consider contracts to contain the snippets that were deployed after its posting.
\item \emph{Source:} A subset of disseminator snippets for which we \textit{only} identified contracts that were deployed after the snippet was posted. These snippets are more likely to have caused SODD.
\end{itemize}

\noindent
Of the 323,328 contracts in our smart contract sanctuary dataset, 135,408 contain code similar to a snippet posted on a Q\&A website. 113,308 contracts show this relation to a (disseminator) snippet that was posted before their deployment. 47,099 contracts show similarity to a (source) snippet that has no embeddings in previously deployed contracts.
To keep all three groups comparable we measure the impact of increased views on increased occurrences by only using snippets with at least one embedding contract (i.e., $\textit{nr} > 0$).
To properly quantify developer adoption of snippets, we measure the correlation between \textit{v} and the number of contracts deployed from unique contract codes containing the snippet \textit{nr}. 
We do not use the Pearson's coefficient as our data is not normally distributed. This could be due to modern search engines recommending highly viewed posts, therefore leading to even more views and more adaptations by developers. We therefore compute the Spearman's rank coefficient $\rho$, a nonparametric correlation used to measure monotonic relationships, i.e., high \textit{v} comes with a high \textit{nr}.

Table~\ref{tab:correlation}, shows our results of computing the coefficient $\rho$ to compare the three groups. No correlation was measured between the views of all snippets and their appearance in contracts. However, with a p-value lower than 0.001 for both disseminator and source snippets, the measured correlation is significantly different from 0. The views \textit{v} of disseminator snippets show only a low correlation of $\rho=0.139$ to the number of newer containing contracts \textit{nr}. For source snippets, the correlation is low to medium with $\rho=0.282$. It is crucial to mention that measuring a correlation is still no prove of a causal relationship. External sources and factors may influence the number of views on a post and the appearance of code in a contract to the same extent. While the correlation is low, we observed several cases of snippets having a high number of views but no identified clones. We therefore suspect that, beside views, other factors play a role in the adoption of snippets. More extensive data crawling and multivariate correlation analysis are necessary to further investigate this issue. However, the presence of a low and low to medium correlation when it comes to disseminator and source snippets supports the decision to investigate them both in the following section.

\begin{table}
    \caption{Spearman correlation $\rho$ of views \textit{v} and number of similar contracts \textit{nr}.}
    \label{tab:correlation}
    \centering
    \begin{tabular}{l | r | r | r }
        \toprule
        \textbf{Temporal Categories} & \textbf{Sample Size} & \textbf{$\rho$} & \textbf{p-value}  \\
        \midrule
        All Snippets         &  4,524 &  0.036 &  0.014 \\
        Disseminator &  3,963 &  0.139 &  <0.001 \\
        Source      &  1,248 &  0.282 &  <0.001 \\ 
        \bottomrule
    \end{tabular}
\end{table}
\subsection{Experiment}
\label{subsec:experiment}

\begin{figure}[t]
\centering
\includegraphics[width=0.9\columnwidth]{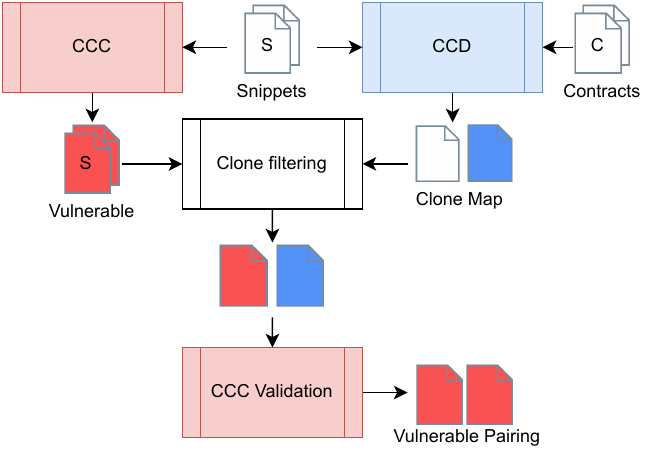}
\caption{Experiment pipeline: 1) snippets are mapped to deployed contracts (\textsc{CCD}), 2) vuln. snippets are identified (\textsc{CCC}), 3) vuln. snippets are validated in contracts (\textsc{CCC}).}
\label{fig:experimentpipeline}
\end{figure}

Figure~\ref{fig:experimentpipeline}, depicts our experiment pipeline that aims to identify whether or not vulnerable code snippets posted on popular Q\&A websites lead to vulnerabilities in Ethereum smart contracts. 
First, we leverage \textsc{CCD} to identify smart contracts that contain clones of the \analyzedsnippets~snippets. 
To produce results with high confidence, we chose conservative parameters: an N-gram size of $3$, an $\eta$ threshold of $0.5$, and an $\epsilon$ threshold of $0.9$.
Next, we run the \analyzedsnippets~snippets through \textsc{CCC}, using the same configuration as in \sectionautorefname{} \ref{subsec:smartbugs}, to identify which snippets are vulnerable.
Afterwards, we map the vulnerable snippets to deployed smart contracts using our mapping produced by \textsc{CCD}.
We restrict our analysis to disseminator and source snippets as defined in Section~\ref{subsec:correlation} to reduce the impact of third-party sources and remove all duplicate smart contracts by comparing the source code after removing comments to account for slight changes that do not alter program behavior. 
This allows us to reduce the cost of the final step and specify how many unique contract implementations are vulnerable.
In our two phased validation, we first run all contracts that were identified to contain vulnerable snippets, through \textsc{CCC}, to validate that the vulnerability is also present in the deployed contract. To avoid counting other vulnerabilities, we run \textsc{CCC} by checking only against the vulnerability that has been previously identified in the vulnerable snippet. In the second phase, we rerun all contract validations that ran into a timeout with modified queries that iteratively reduce the maximal length of data flows. This method finds more true positives by avoiding path explosion problems and does not increase false negatives as the analysis was not terminating previously. This path reduction is only allowed in the second phase of the validation and only in the parts of the query that are not in negated existential subqueries, as they would ignore valid vulnerability mitigations.

\subsection{Results}

Table~\ref{tab:analyzedSamples} shows the result of running our experiment pipeline on the collected snippets and contracts. From the \analyzedsnippets~snippets that were parsable using our modified grammar, we found \vulnsnipp~to be vulnerable. Our conservative configuration of \textsc{CCD} aimed at finding code clones with high confidence, found contracts containing \snippetswclones~ vulnerable snippets. By eliminating contracts that were deployed before the snippet posting we further reduced the amount of snippets to \dissemsnipwc~ that could have led to SODD. This left us with \clonesofdissemsnip~contracts to verify on whether they are vulnerable after including the vulnerable snippets. After filtering duplicate contracts, we remained with \unqclonesofdissemsnip~unique contract clones. \sourcesnipwc~of these potential disseminators of vulnerabilities are source snippets and have \unqclonesofsourcesnip~unique contracts to be verified, shown in parenthesis in Table~\ref{tab:analyzedSamples}.

\begin{table}[t]
    \caption{DASP's Top 10 across snippets and contracts.}
    \label{tab:daspeval}
    \centering
    \begin{tabular}{l | r | r }
        \toprule
        \textbf{Vulnerability Category} & \textbf{Snippets} & \textbf{Contracts} \\
        \midrule
        Reentrancy         &  \snipreentrancy &  \conreentrancy \\
        Denial of Service &  \snipdos &  \condos \\
        Front Running      &  \snipfrontrun &  \confrontrun \\ 
        Time Manipulation  &  \sniptime & \contime \\
        Short Addresses    & \snipshortaddress & \conshortaddress \\
        Access Control & \snipaccess & \conaccess \\
        Arithmetic & \snipover & \conover \\
        Unchecked Low Level Calls & \snipuncheckedcalls & \conuncheckedcalls \\
        Bad Randomness & \snipbadrng & \conbadrng \\
        Unknown Unknowns & \snipunknown & \conunknown \\
        \bottomrule
    \end{tabular}
\end{table}

In the validation step, we run \textsc{CCC} for all \unqclonesofdissemsnip~deployed contracts that are candidates for containing vulnerabilities introduced by using vulnerable code snippets. For the execution we set a timeout of {1,800} seconds per contract and configured the analysis to rerun the vulnerability search that identified a vulnerability in the code snippet. The validation completed for \nrvalidationinfpath~contracts successfully, while it encountered issues such as stack overflows in the Neo4j persistency layer or exceeding the timeout of the analysis for the remaining contracts. With the path reduction method in the second phase, we managed to increase the number of successfully analyzed contracts to \nrvalidationiterpath.
\textsc{CCC} found \vulndissemcontractsiter~of the contracts to be vulnerable and \probdissemsnip~of the \dissemsnipwc~vulnerable snippets to have vulnerable contracts including them. Table~\ref{tab:daspeval} shows the distribution of vulnerability categories across vulnerable snippets and contracts, where multiple snippets and contracts contain more than one vulnerability pattern. It is noteworthy that arithmetic and short address vulnerabilities constitute a large number of validated contracts as they are less relevant for the future. Arithmetic bugs are protected against in versions \texttt{>=0.8}, and addresses have to be sanitized on the client side. Front running issues are mostly low impact but can lead to losses in some cases, and denial of service vulnerabilities can have an impact but are hard to exploit.

\begin{figure}
    \centering
    \fbox{\includegraphics[width=\columnwidth]{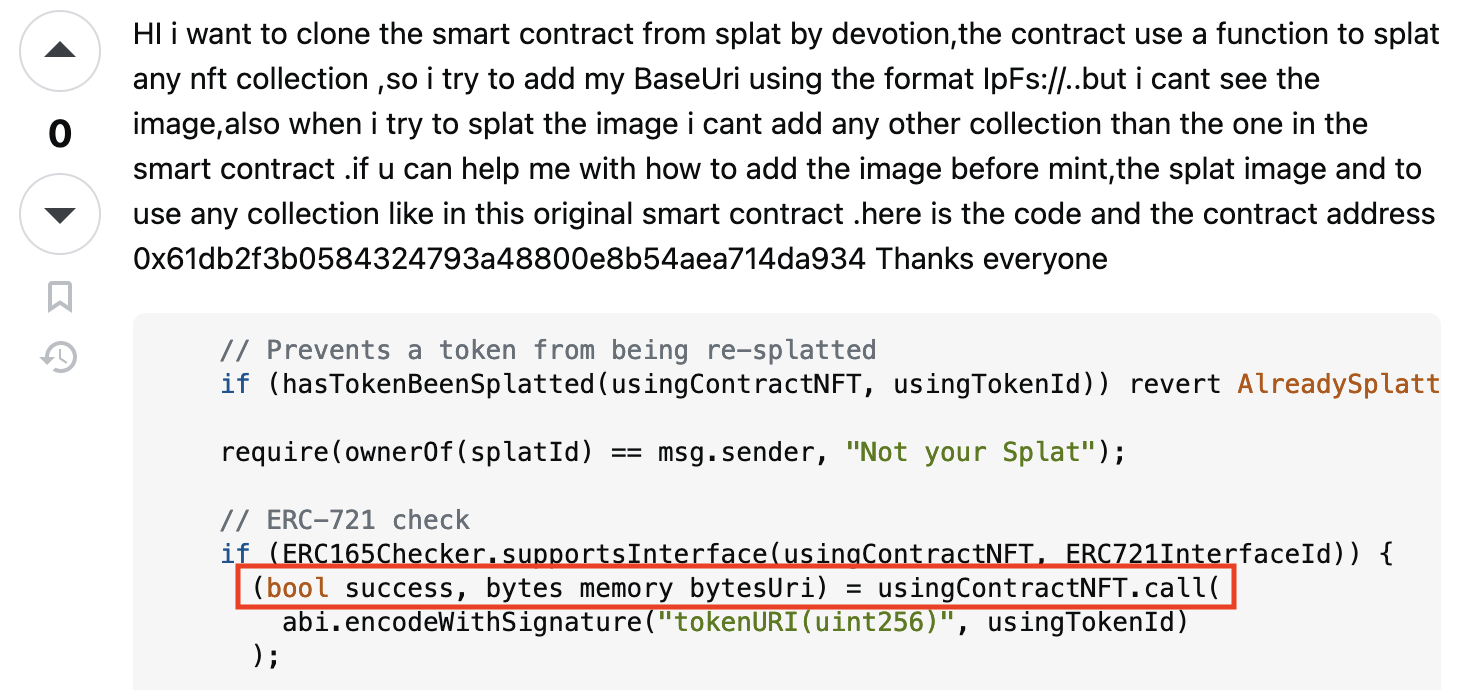}}
    \caption{Example of a snippet on Ethereum Stack Exchange including a potential reentrancy vulnerability.}
    \label{fig:snippet}
\end{figure}

\begin{figure}
    \centering
    \fbox{\includegraphics[width=\columnwidth]{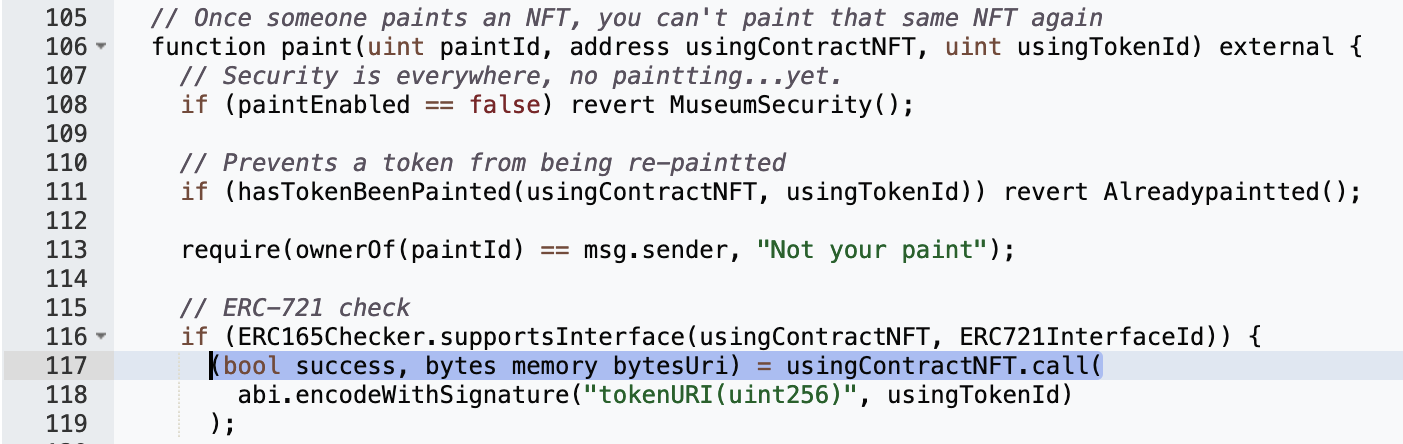}}
    \caption{Example of a deployed smart contract that includes the vulnerable snippet shown in \figureautorefname{} \ref{fig:snippet}.}
    \label{fig:deployed_contract}
\end{figure}

\subsection{Manual Validation}
\noindent
We select \manualsamplesize~contracts that were flagged vulnerable by CCC for our manual analysis. The contracts were equally sampled from all DASP categories when possible, but randomly chosen within the category. This small sample allows us to uncover some qualitative issues of our analysis but does not allow for extrapolation. We manually review: 1) if the original snippet was vulnerable; 2) if the contract was indeed a code clone; 3) if the contract contains the vulnerability. \figureautorefname{} \ref{fig:snippet} depicts a screenshot of a vulnerable snippet containing a potential reentrancy vulnerability posted on the Ethereum Stack Exchange. \figureautorefname{} \ref{fig:deployed_contract} shows a screenshot from Etherscan of a deployed smart contract that includes the vulnerable snippet depicted in \figureautorefname{} \ref{fig:snippet}.
We backtrack every contract to a unique vulnerable snippet they are supposed to be clones of. Both contracts and snippets are unique among each other to create randomized variety. The results of our manual review can be seen in Table~\ref{tab:manualReview}. \manualvcv~of the manually analyzed pairings confirmed to be a vulnerable snippet that had a vulnerable contract identified as a clone. \manualvcn~contracts contained a confirmed vulnerable snippet, but had mitigating circumstance, while in \manualncv~case a vulnerable contract contained a wrongly flagged snippet. In \manualncn~cases \textsc{CCC} incorrectly identified a snippet as vulnerable. In \manualnoclonesum~cases the code clone detection matched a snippet to a contract that we did not consider sufficiently similar. While in \manualvnn~cases the vulnerable snippet had a wrongly identified vulnerability of the same type, in \manualvnv~cases \textsc{CCC} found a vulnerability of the same category that we then confirmed valid. In \manualnnn~cases we flagged snippet and contract incorrectly as vulnerable and clones of each other. In two cases, the snippets were wrongly identified as vulnerable, but the correctly or incorrectly associated contract contained a vulnerable pattern of the same type.

We performed a qualitative analysis of all snippet contract pairings where we either manually identified the snippet or the contract to be a false positive. The result of this qualitative analysis allows us to discuss our limitations when analyzing real-world snippets and contracts. We encountered 37 false positives. This may be due to our small sample size or the unrestricted counting of false positives, see Section~\ref{sec:ccc-exp-results}. We encountered 2 cases of false positives that we have seen in the benchmark analysis. Most other false positives were due to issues we previously did not encounter: 7 FP in the category \texttt{Access Control}, 2 due to complex access controls, one legitimate comparison \texttt{tx.origin != msg.sender}, and 4 cases of proper functionality isolation within the contract initialization; 1 case of a legitimate block number use incorrectly flagged as \texttt{Bad Randomness}; 4 cases of misreported \texttt{Reentrancy} due to different shortcomings of \texttt{CCC}; 6 cases of falsely reported \texttt{Denial of Service} issues where 3 cases were converging loops; 10 cases of harmless patterns to delegate allowances of money transfers being reported as\texttt{Front Running} issues; in 2 cases of \texttt{Arithmetic} issues we did not identify the prevention of the issue through compilation with Solidity 0.8 or greater, and 7 cases where the mitigation for overflows in a SafeMath library was implemented differently than expected. Especially the latter two categories represent some of the more prevalent DASP categories measured in Table~\ref{tab:daspeval}. While extrapolation from such a small sample is not reliable, this could be explained by a high false positive rate over the entire category or the small sample size, requiring further manual investigation. We argue that, even outside of these categories, sufficient snippets and contracts appear to contain vulnerability patterns to raise concerns. In the future, new patterns should be incorporated to account for additional mitigation techniques, e.g., a larger variety of multi-owner, over-underflow checks, and patterns to ignore allowances. Furthermore, graph-based pattern matching should supplemented with techniques to recognize converging loops and unrealistic overflows, e.g., constraint solving and edge case analysis.

\begin{table}
    \centering
    \small
    \caption{Overview of identified vulnerable snippets and contracts across clone matching and validation.}
    \label{tab:analyzedSamples}
    \begin{tabular}{l | r}
        \toprule
        \textbf{Analysis Step} & \textbf{Disseminator} (\textbf{Source}) \\
        \midrule 
        \multicolumn{1}{l}{\textbf{Snippets}} \\
        \midrule
        \hspace{0.2cm} Unique  & \analyzedsnippets \\
        \hspace{0.2cm} Vulnerable & \vulnsnipp \\
        \hspace{0.2cm} Contained in contracts & \snippetswclones \\
        \hspace{0.2cm} Posted before deployment & \dissemsnipwc~(\sourcesnipwc) \\
        \midrule
        \multicolumn{1}{l}{\textbf{Contracts}} \\
        \midrule
        \hspace{0.2cm} Containing vulnerable snippets & \clonesofdissemsnip~(\clonesofsourcesnip) \\
        \hspace{0.2cm} Unique & \unqclonesofdissemsnip~(\unqclonesofsourcesnip) \\
        \midrule
        \multicolumn{1}{l}{\textbf{Validation}} \\
        \midrule
        \hspace{0.2cm} Vulnerable contracts & \vulndissemcontractsiter~(\vulnsourcecontracts)\\
        \hspace{0.2cm} Vuln. snippets in vuln. contracts & \probdissemsnip~(\probsourcesnip)\\
        \bottomrule
    \end{tabular}
\end{table}

\begin{table}
    \centering
    \small
    \caption{Results of our manual validation.}
    \label{tab:manualReview}
    \begin{tabular}{l r | C{2.0cm} C{2.0cm}}
        \toprule
        & \multirow{2}{*}{\textbf{Snippets}} & \multicolumn{2}{c}{\textbf{Contracts}} \\
        & & \textbf{TP} & \textbf{FP} \\
        \midrule
        \multirow{2}{*}{\textbf{True clones}}         & \textbf{TP} & \manualvcv & \manualvcn \\
                                                 & \textbf{FP}   &  \manualncv & \manualncn \\
                                    \midrule
        \multirow{2}{*}{\textbf{False clones}}     & \textbf{TP} &  \manualvnv & \manualvnn \\
                                                 & \textbf{FP}   &  \manualnnv & \manualnnn \\
        \bottomrule
    \end{tabular}
\end{table}

\subsection{Discussion and Limitations}

Our automated analysis identified a concerning number of vulnerable snippets on Q\&A websites. However, many were flagged to be vulnerable to Over-Underflows, Front Running or Short Address attacks. These vulnerabilities depend on the program's context and knowledge of the contract's purpose.
This relation can also be seen within the subset of source snippets. The results of our analysis are not complete as several validations ran into timeouts. We partially mitigated this issue by limiting the length of explored data flow paths for contracts where we received a timeout. This method increased the number of validated patterns in contracts from \vulndissemcontractsinf~to \vulndissemcontractsiter~while not influencing the precision of our results. Only a fraction of the vulnerable snippets has made it into deployed smart contracts. Nonetheless, this small amount of snippets was found in \unqclonesofdissemsnip~contracts, of which \vulndissemcontractsiter~did not mitigate the vulnerability.
\subsection{Mitigations}
In our opinion, the use of snippets from online resources to solve implementation problems cannot be prevented. It is, therefore, necessary to improve the quality and prevalence of pre-deployment static code analysis and steer developers towards more trustworthy sources for examples on how to implement a solution. Providers of Q\&A websites can flag code snippets that are considered problematic by tools like CCC or show high similarity with code reported as part of a vulnerability. Organizations can use static code analysis tools, including tools like CCC, to catch suspicious code fragments in the early development cycle. 
Organizations and companies can provide documentation on the state-of-the-art snippet template to solve frequent problems. The decision on which recommendations are necessary can be made based on the most viewed posts on Q\&A websites. Deviations from whitelisted patterns can be prevented or caught for review in a CI/CD pipeline.

\section{Related Work}
\noindent
In this Section, we present related work on detecting vulnerabilities in smart contracts, identifying code clones, and studying implications of Q\&A websites.

\subsection{Smart Contract Vulnerability Detection}
\noindent
The interest in finding vulnerabilities in smart contracts has been growing in the last few years.  
\textsc{SmartCheck}~\cite{tikhomirov2018smartcheck} translates Solidity source code into an XML-based representation and is checked against XPath patterns. 
\textsc{Slither}~\cite{feist2019slither} uses its own intermediate representation (IR) to transform code into a single static assignment form including data-flow analysis to detect vulnerabilities. 
\textsc{Oyente}~\cite{oyente} uses symbolic execution to find vulnerabilities and was extended~\cite{torres2018osiris} to detect arithmetic vulnerabilities.
\textsc{Mythril}~\cite{mythril} combines symbolic execution with taint analysis to reduce false positives, while \textsc{MAIAN}~\cite{maian} leverages symbolic execution to find vulnerabilities across multiple transactions. 
\textsc{Conkas}~\cite{conkas} also leverages symbolic execution but builds up on \textsc{Rattle}'s IR \cite{rattle}. \textsc{ConFuzzius}~\cite{torres2021confuzzius} presents a hybrid fuzzer that combines symbolic execution with fuzzing. \textsc{HoneyBadger}~\cite{CaminoTBS20} uses symbolic execution to identify smart contract scams. 
\noindent
Sendner et al.~\cite{sendner} performed a large-scale analysis of several vulnerability detection tools and found the state-of-the-art insufficient in tackling the challenge of detecting vulnerabilities.
\noindent
Giesen et al.~\cite{giesen2022practical} developed a compiler for smart contracts (HCC) which translates smart contracts into a CPG to patch reentrancy and integer bugs. While the underlying concept of using a CPG is similar, \textsc{CCC}'s CPG has been developed with to analyze code snippets an full smart contracts. In fact, the aforementioned tools either only operate on bytecode or are not able to analyze incomplete code such as snippets.

\subsection{Code Clone Detection}
\noindent
A plethora of tools have been proposed to detect smart contract code clones at bytecode or source code level. 
\textsc{EClone}~\cite{liu2018eclone} compares contracts based on their symbolic transaction sketches and extended their approach by introducing birthmarking~\cite{LiuYJZS19}. 
\textsc{Eth2Vec}~\cite{ashizawa2021eth2vec} uses natural language processing on Ethereum bytecode to argue on the similarity of code.
\noindent
Zhu et al.~\cite{electronics11040597} uses neural networks to detect similar bytecode while being resilient to different compiler versions and later used pattern matching over control-flow graphs to detect similarity \cite{zhuCFG2021}. 
\textsc{Deckard}~\cite{jiang2007deckard} presents an efficient algorithm to cluster syntax sub-trees of C and Java, and was adapted for Solidity. \textsc{SmartEmbed}~\cite{gao2019smartembed} uses structural code embeddings to detect code clones in Solidity source code. White et al.~\cite{WhiteTVP16} used a deep learning approach to detect code clones with higher accuracy than \textsc{Deckard}. 
Kondo et al.~\cite{KondoOJHM20} used \textsc{Deckard} to identify code clones among contracts listed on Etherscan and compared if they contained code blocks from OpenZeppelin \cite{openzeppelin}. 
Chen et al.~\cite{ChenLZ0Z21} studied code reuse among contracts on Etherscan. Similar to \textsc{CCD}, they parse Solidity with ANTLR and tokenize the AST, but compare on a contract level, the contract name and a rigid token sequence. 
\textsc{Volcano}~\cite{fatima2022volcano} uses the tool \textsc{NiCad} \cite{roy2008nicad} to find vulnerable smart contracts by comparing their code on a function level with known vulnerable contracts.
Most aforementioned tools are either designed to detect bytecode similarity or are not specifically designed to detect clones of code snippets, making them unable to parse code snippets.
Similar to us, He et al.~\cite{he2020clones} uses fuzzy hashing to compare smart contracts. However, the authors analyze bytecode, while we compare source code and thus normalize and tokenize with a different techniques. Moreover, we only preserve code order for atomic code blocks, and our approach is more efficient due to the pre-filtering via n-grams.

\subsection{Implications of Q\&A Websites}

Fischer et al.~\cite{fischer2017stack} presented the first automated approach to detect security-relevant code snippets from a Q\&A website in Android Applications. Their approach uses machine learning based classification to get a security score and abstract interpretation on program dependence graphs to identify the presence of snippets.
Ayman et al.~\cite{afia2019} studied discussions related to Ethereum smart contracts posted on Stack Overflow but did not analyze security implications of vulnerable code snippets. 
Soud et al.~\cite{soud2022} conducted an empirical study on smart contract related vulnerabilities found on Stack Overflow and GitHub, but they performed a manual analysis and did not evaluate the impact on deployed contracts. 
Our work, is the first to analyze the transfer of vulnerable code snippets to real-world smart contracts in a fully automated fashion.

\section{Conclusion}
\noindent
This paper investigates the introduction of vulnerable code through copy-pasting snippets from Q\&A websites to deployed Solidity smart contracts.
We validate our vulnerability detection \textsc{CCC} and our code clone detection \textsc{CCD} by comparing them to other tools in their field. Our results show that both are competitive in their field and even outperform their counterparts. The most important requirement that these tools fulfill and prior tools did not is the ability to analyze incomplete code, which is essential for analyzing code snippets posted on Q\&A websites and whether they are included in deployed smart contacts.

We found \vulnsnipp~vulnerable code snippets across Q\&A websites. Using our clone detection, we identified \unqclonesofdissemsnip~contracts with unique code that contains fragments related to \dissemsnipwc~vulnerable code snippets from Stack Overflow and Ethereum Stack Exchange. \vulndissemcontractsiter~of the \nrvalidationiterpath~successfully analyzed contracts were flagged as vulnerable as they did not implement mitigation measures. 
While direct causality cannot be measured, as mentioned in Section~\ref{subsec:conceptuallimitaitons}, we consider our filtering of code clones based on timestamps and categorizing into dissemination and source snippets to sufficiently support the evidence that the problem of copy-paste code exists in the Ethereum ecosystem. 

In the future, we will extend the number of vulnerability searches and analyze them on a larger scale by solving scalability issues of the graph database and path explosion in large smart contracts. Iteratively reducing the length of our data flow paths will allow us to confirm even more vulnerabilities in deployed smart contracts without negatively influencing the accuracy of the validation. 

\begin{acks}
This work was supported by the German Federal
Ministry of Education and Research (BMBF) project 6G-ANNA
(16KISK087) and the Zurich Information Security \& Privacy Center (ZISC).
\end{acks}

\bibliographystyle{ACM-Reference-Format}
\balance
\bibliography{references}

\appendix

\section{Ethics}
We discussed and disclosed security issues in accordance with our Offensive and Ethics Review Board. The vulnerability patterns and evaluation tools presented in this work are not considered exploits and neither enable non-specialist attackers nor increase the efficiency of professional attackers. The number of analyzed contracts, as well as the anonymity of contract owners makes it difficult to seek contact and report our findings.

\section{Vulnerability Queries}
\label{sec:appendix}

\begin{lstlisting}[language=Cypher,caption={Access Control: Unrestricted writes to state variable used for access control.},captionpos=b, ,label={lst:accesscontrollogic}, basicstyle=\scriptsize]
match p=(entry:FunctionDeclaration)-[e:EOG|INVOKES|RETURNS*]->(wN)-[:EOG|INVOKES|RETURNS*]->(last)
where not 'ConstructorDeclaration' in labels(entry) and not split(entry.code,'{')[0] contains ' internal ' and not exists((last)-[:EOG|INVOKES]->())
 and exists((wN)-[:DFG]->(:FieldDeclaration)<-[:REFERS_TO]-()<-[:LHS|RHS]-(:BinaryOperator {operatorCode:'=='})-[:LHS|RHS]->({code: 'msg.sender'}))
 and not exists{
  match ({code:'msg.sender'})-[:DFG*]->(n)<-[:DFG*]-(:FieldDeclaration)
  match alt=(n)-[:DFG*]->(comp)-[:EOG|INVOKES|RETURNS*]->(t)
  where comp in nodes(p)
  and ('Rollback' in labels(t) or not wN in nodes(alt))
} return entry
\end{lstlisting}
\begin{lstlisting}[language=Cypher,caption={Access Control: Unrestricted access to functions that destroy the smart contract.},captionpos=b, ,label={lst:accesscontrolselfdestruct}, basicstyle=\scriptsize]
match p=(f:FunctionDeclaration)-[:EOG|INVOKES*]->(c:CallExpression)-[:EOG|INVOKES*]->(last)
 where toUpper(c.localName) in ['SELFDESTRUCT','SUICIDE']
 and not exists ((last)-[:EOG|INVOKES]->())
 and not 'Rollback' in labels(last)
 and not exists{
  ({code : 'msg.sender'})-[:DFG*]->(n)-[:EOG|INVOKES*]->(t)
  where n in nodes(p) and not exists ((t)-[:EOG|INVOKES]->())
  and exists{
   alt=(f)-[:EOG|INVOKES*]->(n)-[:EOG|INVOKES*]->(t) where 'Rollback' in labels(t) or not c in nodes(alt)
}} return c
\end{lstlisting}
\begin{lstlisting}[language=Cypher,caption={Short Addresses: Address padding issues at callsites in the contract.},captionpos=b, ,label={lst:addresspaddingatcall}, basicstyle=\scriptsize]
match p=({localName:"address"})<-[:TYPE]-(ad)<-[adr:PARAMETERS]-(f:FunctionDeclaration)-[:EOG|INVOKES*]->(c:CallExpression)-[:EOG|INVOKES*]->(last)
 where ('ReturnStatement' in labels(last) or exists {(f)-[:BODY]->(last)})
  and not split(f.code, '{')[0] contains ' internal '
  and (toUpper(c.localName) in ['TRANSFER' , 'SEND']
  and exists{
   (f)-[r:PARAMETERS]->(param:ParamVariableDeclaration)-[:DFG*]->()<-[:ARGUMENTS]-(c)
   where not exists {(f)-[rp:PARAMETERS]->() where rp.INDEX > r.INDEX} and adr.INDEX < r.INDEX
   }
  or exists{
   (f)-[r:PARAMETERS]->(param:ParamVariableDeclaration)-[:DFG*]->()<-[:VALUE]-(s)-[:KEY]->({value: 'value'})
   where exists{(s)<--(c)} and not exists {(f)-[rp:PARAMETERS]->() where rp.INDEX > r.INDEX} and adr.INDEX < r.INDEX
  }
  or toUpper(c.localName) in ['VALUE'] and exists{
   (f)-[r:PARAMETERS]->(param:ParamVariableDeclaration)-[:DFG*]->()<-[:ARGUMENTS]-(c)-[:BASE|CALLEE*]->({localName:'call'})
   where not exists {(f)-[rp:PARAMETERS]->() where rp.INDEX > r.INDEX} and adr.INDEX < r.INDEX
  })
 and not exists{
  ({code : 'msg.data.length'})-[:DFG*]->(n)
  where n in nodes(p)
  and exists{alt=(n)-[:EOG|INVOKES*]->(t) where 'ROLLBACK' in labels(t) or not c in nodes(alt) and not exists {(t)-[:EOG|INVOKES]->()}}
 } and exists{
   (c)-[:BASE|CALLEE*]->()<-[:DFG*]-(:ParamVariableDeclaration)
} return c
\end{lstlisting}
\begin{lstlisting}[language=Cypher,caption={Short Addresses: Writes to contract state vulnerable to address padding attacks.},captionpos=b, ,label={lst:addresspaddingatstate}, basicstyle=\scriptsize]
match p=({localName:"address"})<-[:TYPE]-(ad)<-[adr:PARAMETERS]-(f:FunctionDeclaration)-[:EOG|INVOKES*]->(last)
 where ('ReturnStatement' in labels(last) or exists {(f)-[:BODY]->(last)}) 
  and exists{
  (f)-[vulna:PARAMETERS]->(vuln)-[:DFG*]->(m)-[:DFG*]->(state:FieldDeclaration)
   where not exists {(f)-[rp:PARAMETERS]->() where rp.INDEX > vulna.INDEX} and adr.INDEX < vulna.INDEX
   and not exists{
    ({code:'msg.data.length'})-[:DFG*]->(n)
    where n in nodes(p)
    and exists{alt=(n)-[:EOG|INVOKES*]->(t) where 'ROLLBACK' in labels(t) or not m in nodes(alt) and not exists {(t)-[:EOG|INVOKES]->()}}
 }    
} return ad
\end{lstlisting}
\begin{lstlisting}[language=Cypher,caption={Bad Randomness: Usages of bad sources for randomness.},captionpos=b, ,label={lst:badrandomness}, basicstyle=\scriptsize]
match (r) where ((r:DeclaredReferenceExpression or r:MemberExpression)  and r.code IN ["block.timestamp", "block.number", "block.difficulty", "block.coinbase"]
or (r:CallExpression and r.localName in ['blockhash']))
and ( exists {
 (r)-[:DFG*]->(:ReturnStatement)<-[:EOG*]-(containing:FunctionDeclaration) where containing.code contains 'rand'
} or exists {
  (r)-[:DFG|ARGUMENTS*]->(f:FieldDeclaration)
  where not exists((f)-[:DFG]->())
} or exists{
 (int:CallExpression)
 where int.localName in ['value', 'send', 'transfer', 'call'] and (
  exists{
   (r)-[:DFG*]->()<-[:BASE|CALLEE|ARGUMENTS|SPECIFIERS|VALUE*]-(int)
  } or exists {
   (r)-[:DFG*]->(branch)-[:EOG]->(th)-[:EOG*]->(int) where exists {(branch)-[:EOG]->(el) where el <> th
   and (int:Rollback or int:CallExpression)
   and not exists ((el)-[:EOG*]->(int))}
})}) return r
\end{lstlisting}
\begin{lstlisting}[language=Cypher,caption={Denial of Service: External call whoes failure prevents execution of other money transfering calls.},captionpos=b, ,label={lst:callblocksfuturecall}, basicstyle=\scriptsize]
match (c:CallExpression)-[:EOG*]->(c2:CallExpression)
where c.localName in ['transfer', 'send', 'call'] and c2.localName in ['transfer','send', 'call']
and (not c.localName in ['transfer', 'send'] or exists{
 avoidingpath=(c)-[:DFG]->(branchNeg)-[:EOG]->(next) where not exists ((next)-[:EOG*]->(c2))
}) return c
\end{lstlisting}
\begin{lstlisting}[language=Cypher,caption={Denial of Service: External call whoes failure prevents state changes.},captionpos=b, ,label={lst:callblocksfuturestatechanges}, basicstyle=\scriptsize]
match (c:CallExpression)-[:EOG*]->(write1)-[:DFG]->(f:FieldDeclaration)
where (c.localName in ['transfer'] or c.localName = 'send' and exists{
 avoidingpath=(c)-[:DFG]->(branchNeg)-[:EOG*]->(last) where not exists ((last)-[:EOG]->()) and not write1 in nodes(avoidingpath)
}) and not exists { alt=(f)<-[:DFG]-(write2)-[:EOG*]->(func:FunctionDeclaration)
 where not 'ConstructorDeclaration' in labels(f) and not c in nodes(alt) and not exists {
  (write2)-[:EOG*]->(branching)-[:EOG*]->(c)
 }
} return c
\end{lstlisting}
\begin{lstlisting}[language=Cypher,caption={Unchecked Low Level Calls: Critical calls were return values are ignored.},captionpos=b, ,label={lst:uncheckedreturn}, basicstyle=\scriptsize]
match p=(c:CallExpression)-[:EOG*]->(last)
where not exists ((last)-[:EOG]->()) and not 'Rollback' in labels(last)
 and not exists{(c)-[:DFG*]->(r:ReturnStatement) where r in nodes(p)}
 and not exists{
  (c)-[:DFG*]->(n)-[:EOG]->(apath) where n in nodes(p) and exists {
   (n)-[:EOG]->(otherpath) where apath <> otherpath
  }
}and (c.localName in ['call', 'callcode', 'delegatecall', 'send']
 or c.localName in ['value','gas'] and exists {
  (c)-[:BASE|CALLEE*]->({localName:'call'})
 }
) return c
\end{lstlisting}
\begin{lstlisting}[language=Cypher,caption={Denial of Service: Expensive loops that can be used by an attacker to consume large quantities of gas.},captionpos=b, ,label={lst:dosthroughExhaustion}, basicstyle=\scriptsize]
match p=(b)-[:EOG*]->(cond)-[:EOG]->(b)
where (b:ForStatement or b:WhileStatement or b:DoStatement or b:ForEachStatement)  and (exists {(exp)-[DFG]->(:FieldDeclaration) where exp in nodes(p)}
 or exists{(exp:CallExpression) where exp in nodes(p) and not exists ((exp)-[:INVOKES]->()) or exists {(exp)-[:INVOKES]->(target) where not exists {(target)-[:BODY]->()}}}
 ) and  (
 exists {(l:Literal)-[:DFG]->(cond:BinaryOperator) where cond.operatorCode in ['<','<=','>','>='] and l.value > 100}
 or exists { (cond)<-[:DFG*]-(userC:ParamVariableDeclaration)<-[:PARAMETERS]-(f:FunctionDeclaration) where not 'ConstructorDeclaration' in labels(f)}
 ) return b
\end{lstlisting}
\begin{lstlisting}[language=Cypher,caption={Access Control: Call delegation vulnerabilities where inputs are not properly sanitized.},captionpos=b, ,label={lst:defaultProxyDelegate}, basicstyle=\scriptsize]
match p=(f:FunctionDeclaration)-[:EOG|INVOKES*]->(c:CallExpression)-[:EOG|INVOKES*]->(last)
where (f.localName IS NULL or f.localName = null or f.localName = '') and  toUpper(c.localName) in ['DELEGATECALL' , 'CALLCODE']
and not exists ((last)-[:EOG|INVOKES]->()) and not 'Rollback' in labels(last)
and (exists {({code: 'msg.data'})<-[:ARGUMENTS]-(c)
} or exists{({code: 'msg.data'})-[:DFG*]->()<-[:ARGUMENTS]-(c)})
and not exists{
 df=(source {code: 'msg.data'})-[:DFG*]->(n)-[:EOG]->(apath) where n in nodes(p)
 and not exists {(otherf:FunctionDeclaration|CallExpression) where otherf in nodes(df)}
 and not exists ((source)<-[:BASE]-({code: 'msg.data.length'}))
 and exists {
 d=(f)-[:EOG|INVOKES*]->(n)-[:EOG|INVOKES*]->(otherpath) where not exists{ (otherpath)-[:EOG|INVOKES]->() } and (not c in nodes(d) or 'Rollback' in labels(otherpath))
}} return c
\end{lstlisting}
\begin{lstlisting}[language=Cypher,caption={Denial of Service: Collections that are used for transfers and can be cleared outside of contract initialization.},captionpos=b, ,label={lst:emptycollectiontransfer}, basicstyle=\scriptsize]
match p=(b:BinaryOperator {operatorCode: '='})-[:LHS]->()-[:DFG]->(state:FieldDeclaration)-[:TYPE]->(t)
where t.code contains '[' and exists{
 (c:CallExpression)-[:BASE|CALLEE|ARGUMENTS]->()<-[:DFG*]-(state) where c.localName in ['transfer', 'send', 'call']
} and not exists {(f:ConstructorDeclaration)-[:EOG*]->(b)}
return b
\end{lstlisting}
\begin{lstlisting}[language=Cypher,caption={Front Running: Code where a miner can gain the same beneficial state change as any other transaction sender would.},captionpos=b, ,label={lst:frontrunning}, basicstyle=\scriptsize]
match p=(f:FunctionDeclaration)-[:EOG*]->(int)-[:EOG*]->(last)
where not 'ConstructorDeclaration' in labels(f) and not exists((last)-[:EOG]->())
and (exists{ (int:BinaryOperator {operatorCode: '='})-[:LHS]->()<-[:DFG*]-(sourcer {code: 'msg.sender'})
 where not exists{(int:BinaryOperator)-[:RHS]->(rhs)<-[:DFG*]-(source) where source.code = "msg.sender" or source.code = "msg.value"}
} or exists {
(int:CallExpression)-[:BASE|CALLEE*]->(target {code: 'msg.sender'})
where int.localName in ['value', 'send', 'transfer', 'call']
and not exists {({code: 'msg.sender'})-[:DFG*]->()<-[:ARGUMENTS]-(int)}
 or exists {(int)-[:BASE|CALLEE*]->(SpecifiedExpression)-[:SPECIFIERS]->(kv:KeyValueExpression)-[:KEY]->({localName:'value'})
  where not exists {({code: 'msg.sender'})-[:DFG*]->()<-[:VALUE]-(kv)}
}}) and not exists {
 match alt=(f)-[:EOG*]->(branch)-[:EOG*]->(altlast)
 match (source {code: 'msg.sender'})-[:DFG*]->(branch)
 where not exists((altlast)-[:EOG]->()) and branch in nodes(p) and source in nodes(p) and (not int in nodes(alt) or altlast:Rollback)
} return int
\end{lstlisting}
\begin{lstlisting}[language=Cypher,caption={Unknown Unknowns: Writes to local structs that can lead to unintentional overwriting of state variables.},captionpos=b, ,label={lst:localwritetostorage}, basicstyle=\scriptsize]
match (v:VariableDeclaration)
where ('ParamVariableDeclaration' in labels(v) and ' storage ' in v.code
or not 'ParamVariableDeclaration' in labels(v) and not 'FieldDeclaration' in labels(v) and not exists{(dc)-[:AST]-(v) where dc.code contains ' memory ' or dc.code contains ' calldata '}) and not exists ((v)-[:INITIALIZER]->())
and ('[' in v.code or exists{ 
  (v)-[:TYPE]->(tv) where exists{
  (struct:RecordDeclaration {kind: 'struct'}) where struct.kind = 'struct' and struct.localName = tv.localName
 }}
) and exists {(f) where not 'ConstructorDeclaration' in labels(f) and (
 exists{(f)-[:EOG*]->()-[d:DFG]->(v)
 } or exists {
  (f)-[:EOG*]->()-[d:DFG]->(bin:BinaryOperator)-[:LHS]->()-[:BASE|CALLEE|LHS|ARRAY_EXPRESSION*]->()<-[:DFG*]-(v)
  where bin.operatorCode in ['=', '|=', '^=', '&=', '<<=','>>=','+=', '-=', '*=', '/=', '%=']
 } or exists {
  (f)-[:EOG*]->()-[d:DFG]->(bin:UnaryOperator)-[:INPUT|BASE|CALLEE|LHS|ARRAY_EXPRESSION]->()<-[:DFG*]-(v)
  where bin.operatorCode in ['++','--']
})} return v
\end{lstlisting}
\begin{lstlisting}[language=Cypher,caption={Arithmetic: Arithmetic operations that can over- or underflow.},captionpos=b, ,label={lst:overunderflow}, basicstyle=\scriptsize]
match p=(f:FunctionDeclaration)-[:EOG*]->(b:BinaryOperator)-[:EOG*]->(last)
where not exists((last)-[:EOG*]->()) and b.operatorCode in ['+','+=','-','-=','*','*=']
and exists {(b)<-[:DFG*]-(param:ParamVariableDeclaration)<--(argf:FunctionDeclaration) where not 'ConstructorDeclaration' in labels(f) and not split(argf.code, '{')[0] contains ' internal '}
and((exists{
  (b)-[d:DFG*]->(:FieldDeclaration)
 } or exists {
  (b)-[d:DFG*]->(bin:BinaryOperator)-[:DFG]->()-[:EOG]->(:Rollback)
   where bin.operatorCode in ['<', '>', '<=', '>=', '==']
 } or exists {
  (b)-[d:DFG*]->(bin:BinaryOperator)-[:LHS]->()-[:BASE|CALLEE|LHS|ARRAY_EXPRESSION*]->()<-[:DFG*]-(:FieldDeclaration)
  where bin.operatorCode in ['=', '|=', '^=', '&=', '<<=','>>=','+=', '-=', '*=', '/=', '%=']
 }or exists {
  (b)-[d:DFG*]->(bin:UnaryOperator)-[:INPUT|BASE|CALLEE|LHS|ARRAY_EXPRESSION]->()<-[:DFG*]-(:FieldDeclaration)
  where bin.operatorCode in ['++','--']
 } or exists{
  (b)-[:DFG*]->()<-[:ARGUMENTS]-(c:CallExpression) where not exists((c)-[:INVOKES]->()-[:BODY]->())
 } or exists{
  (b)<-[:ARGUMENTS]-(c:CallExpression) where not exists((c)-[:INVOKES]->()-[:BODY]->())
 }or exists {
  (b)-[:DFG*]->()<-[:VALUE]-(:SpecifiedExpression)
 }or exists {
  (b)<-[:VALUE]-(:SpecifiedExpression)
}) and not exists {
  match bpath=(f)-[:EOG*]->(cond:BinaryOperator)-[:EOG]->(branch)-[:EOG*]->(l)
  match (c1)<-[:LHS|RHS]-(cond)-[:LHS|RHS]->(c2)
  where c1 <> c2 and branch in nodes(p) and not exists((l)-[:EOG]->())
  and (not b in nodes(bpath) or 'Rollback' in labels(l))
  and not exists {
   (dfOrigin)-[:DFG*]->(b) where not exists(()-[:DFG]->(dfOrigin)) and not exists ((dfOrigin)-[:DFG*]->(branch))
  } and (not exists{(b)-[:DFG*]->(branch)} or
   exists ((b)<-[:DFG*]-()-[:DFG*]->(c1))
   and exists ((b)<-[:DFG*]-()-[:DFG*]->(c2))
   or exists ((:Literal)-[:DFG]->(cond)) and exists ((:Literal)-[:DFG]->(b))
)}) return b
\end{lstlisting}
\begin{lstlisting}[language=Cypher,caption={Reentrancy: Callpaths through external calls vulnerable to reentrancy attacks.},captionpos=b, ,label={lst:reentrancy}, basicstyle=\scriptsize]
match p=(base:MemberExpression)-[:BASE|CALLEE]-(c:CallExpression)-[e:EOG|INVOKES|RETURNS*]->(n)
where not exists {(c)<--(em:EmitStatement)}
and not exists{
 ()-[r:RETURNS]->()-[i:INVOKES]->()
 where r in relationships(p) and apoc.coll.indexOf(relationships(p), r) + 1 = apoc.coll.indexOf(relationships(p), i)
}and(exists{
  (n)-[d:DFG*]->(field:FieldDeclaration)
  where exists ((field)<-[:FIELDS]-(:RecordDeclaration)-[:AST*]->(c))
 } or exists {
  (n)-[d:DFG*]->(bin:BinaryOperator)-[:LHS]->()-[:BASE|CALLEE|LHS|ARRAY_EXPRESSION*]->()<-[:DFG*]-(field:FieldDeclaration)
  where bin.operatorCode in ['=', '|=', '^=', '&=', '<<=','>>=','+=', '-=', '*=', '/=', '%=']
  and exists ((field)<-[:FIELDS]-(:RecordDeclaration)-[:AST*]->(c))
 }or exists {
  (n)-[d:DFG*]->(bin:UnaryOperator)-[:INPUT|BASE|CALLEE|LHS|ARRAY_EXPRESSION]->()<-[:DFG*]-(field:FieldDeclaration)
  where bin.operatorCode in ['++','--']
  and exists ((field)<-[:FIELDS]-(:RecordDeclaration)-[:AST*]->(c))
 })and(not exists  {()-[:DFG]->(b1)<-[:BASE|CALLEE*]-(c)}
or exists {
 dflow=(s)-[:DFG*]->(b2)<-[:BASE]-(callee)<-[:CALLEE]-(c)
 where(exists((b2)-[:TYPE]->({name: "address"})) or exists((b2)-[:TYPE]->(:ObjectType)-[:RECORD_DECLARATION]->()))
 and not exists (()-[:DFG]->(s)) and  not 'Literal' in labels(s) and not  exists((s)<-[:PARAMETERS]-(:ConstructorDeclaration)) and (not s.isInferred or s.localName in ['msg', 'tx'] )
 and not exists{(sub)-[:DFG]->(array)-[:SUBSCRIPT_EXPRESSION]->(sub) where sub in nodes(dflow) and array in nodes(dflow)}
}) and (
 exists{((d:DeclaredReferenceExpression)-[:DFG*]->(base)) where d.code in ['msg.sender', 'tx.origin']}
 or exists{((t {localName: "address"})<-[:TYPE]-(root)-[:DFG*]->(base)) where t.localName ='address' or t.localName = 'UNKNOWN' and not exists((root)<-[:DFG]-()) }
)return c
\end{lstlisting}
\begin{lstlisting}[language=Cypher,caption={Time Manipulation: Transactions where a miner can choose the time of execution to change the outcome.},captionpos=b, ,label={lst:timemanipulation}, basicstyle=\scriptsize]
match (r:DeclaredReferenceExpression)
where r.code in ['now','block.timestamp']
and (exists { (r)-[:DFG*]->(:ReturnStatement)}
or exists{
 (r)-[:DFG*]->(exp:CallExpression) where not exists ((exp)-[:INVOKES]->()) or exists {(exp)-[:INVOKES]->(target) where not exists {(target)-[:BODY]->()}}
} or exists {
 (r)-[:DFG*]->(:FieldDeclaration)
} or exists {
 (r)-[:DFG*]->(branch)-[:EOG]->(th)-[:EOG*]->(int) where exists {(branch)-[:EOG]->(el) where el <> th
 and (int:Rollback or int:CallExpression)
 and not exists ((el)-[:EOG*]->(int))}
}) return r
\end{lstlisting}
\begin{lstlisting}[language=Cypher,caption={Access Control: Uses of tx.origin for branching.},captionpos=b, ,label={lst:txorigin}, basicstyle=\scriptsize]
match (:FieldDeclaration)<-[:REFERS_TO]-()-[:DFG*]->(n)
match (:MemberExpression {code : 'tx.origin'})-[:DFG*]->(n)
match (b1)<-[:EOG]-(n)-[:EOG]->(b2)
where b1 <> b2
return n
\end{lstlisting}

\section{CCD Parameters}
\label{sec:appendix_code_clone_detection_parameters}

\tableautorefname{} \ref{tab:parameters}, lists the different combinations of N-gram sizes, N-gram thresholds, and similarity thresholds that we tried in our comparison with our clone detection tool against \textsc{SmartEmbed}.

\begin{table}[H]
    \centering
    \small
    \begin{tabular}{c l l}
        \toprule
        \textbf{Parameter} & \textbf{Description} & \textbf{Values} \\
        \midrule
        $N$ & N-gram size & $3, 5, 7$ \\
        $\eta$ & N-gram threshold & $0.5, 0.6, 0.7, 0.8, 0.9$ \\
        $\epsilon$ & Similarity threshold & $0.5, 0.6, 0.7, 0.8, 0.9$ \\
        \bottomrule
    \end{tabular}
    \caption{List of parameters used in our experiments for our code clone detection tool.}
    \label{tab:parameters}
\end{table}

\section{CCD Parameter Comparison}
\label{sec:appendix_code_clone_detection_parameter_results}
\begin{figure*}
     \centering
     \begin{subfigure}[b]{0.32\textwidth}
         \centering
         \includegraphics[width=\textwidth]{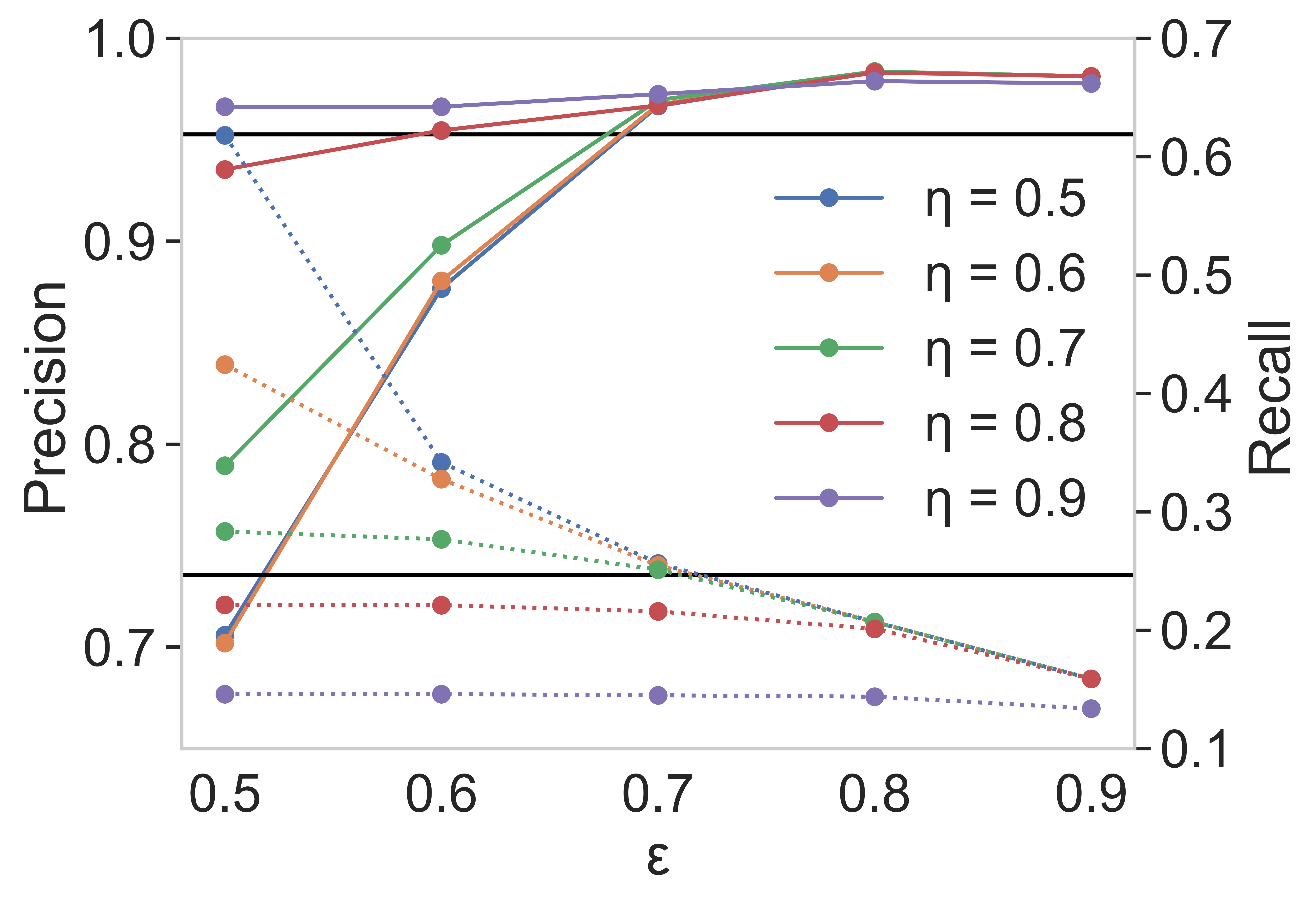}
         \caption{N-gram size 3}
         \label{fig:ngram_size_3}
     \end{subfigure}
     \hfill
     \begin{subfigure}[b]{0.32\textwidth}
         \centering
         \includegraphics[width=\textwidth]{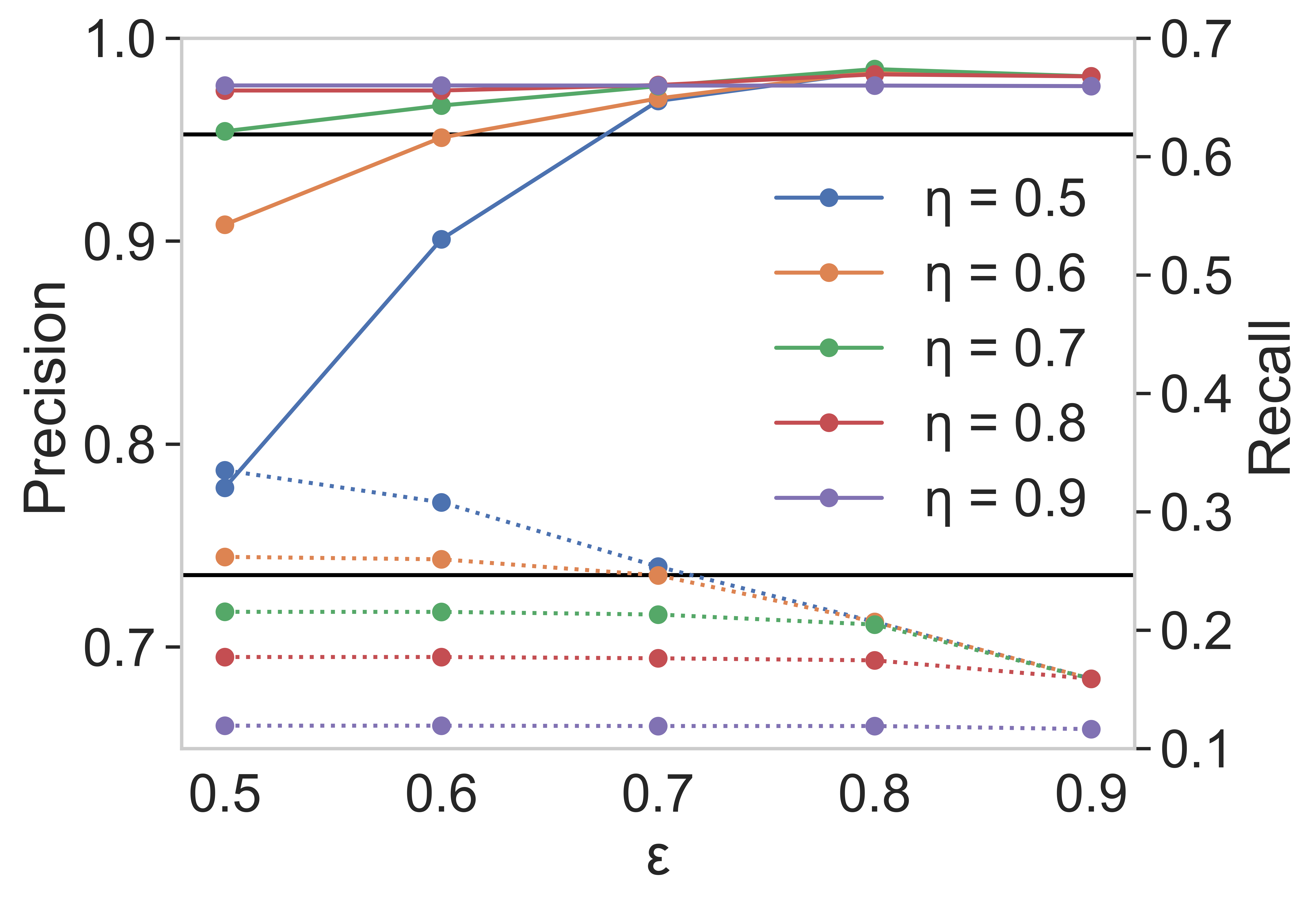}
         \caption{N-gram size 5}
         \label{fig:ngram_size_5}
     \end{subfigure}
     \hfill
     \begin{subfigure}[b]{0.32\textwidth}
         \centering
         \includegraphics[width=\textwidth]{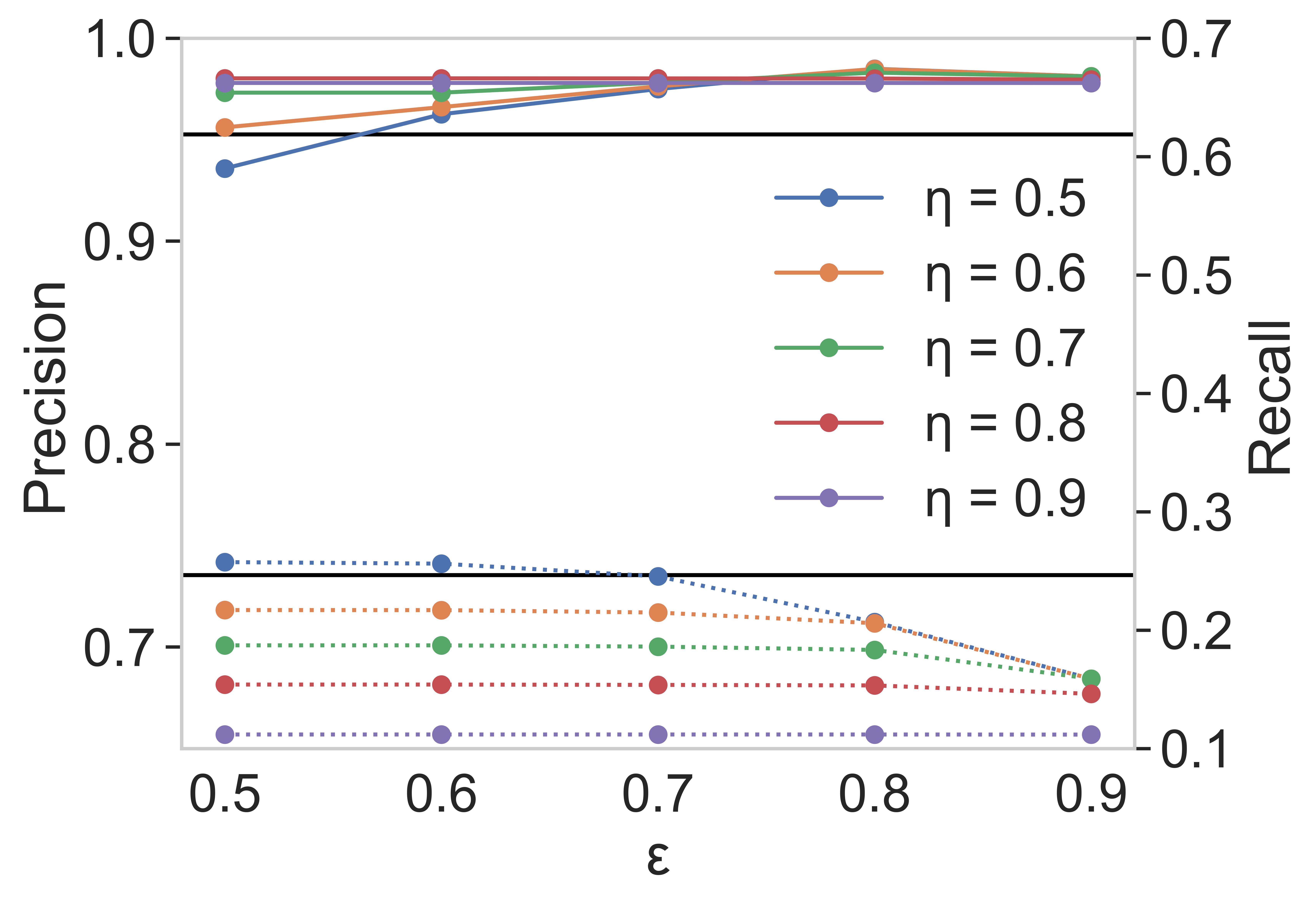}
         \caption{N-gram size 7}
         \label{fig:ngram_size_7}
     \end{subfigure}
        \caption{Precision and recall comparison for different N-gram sizes, $\eta$ thresholds, and $\epsilon$ thresholds.}
        \label{fig:parameter_results}
\end{figure*}
\figureautorefname{} \ref{fig:parameter_results}, shows the precision (solid lines) and recall (dashed lines) achieved by our code clone detection across different combinations of parameters values as defined in \tableautorefname{} \ref{tab:parameters}.
The upper black horizontal line defines the precision achieved by \textsc{SmartEmbed}, whereas the lower one defines the recall achieved by \textsc{SmartEmbed}. Our goal is to find a combination of parameters that results in a precision that is above the upper black horizontal line and a recall that is above the lower black horizontal line.
Our code clone detection tool achieved the highest precision (i.e., $0.98491$) using an N-gram size of $7$, an $\eta$ threshold of $0.5$, and an $\epsilon$ threshold of $0.8$, but the recall was as expected, very low (i.e., $0.2068$).
The highest recall (i.e., $0.617966$) was achieved using an N-gram size of $3$, an $\eta$ threshold of $0.5$, and an $\epsilon$ threshold of $0.5$, but the precision was on the other hand very low (i.e., $0.617966$).
The best combination of both precision (i.e., $0.966586$) and recall (i.e., $0.256293$) was achieved using an N-gram size of $3$, an $\eta$ threshold of $0.5$, and an $\epsilon$ threshold of $0.7$.

\end{document}